\documentclass[final,5p,times,twocolumn]{elsarticle}
\usepackage{amssymb}
\usepackage{amsmath}
\usepackage{lineno}

\usepackage{setspace}

\usepackage{bm}
\usepackage[colorinlistoftodos]{todonotes}
\usepackage{verbatim}
\usepackage{epsfig}
\usepackage[export]{adjustbox}
\usepackage{moresize}

\newcommand{\Vector}[1]{\ensuremath{\mathbf{#1}}}
\newcommand{\Tensor}[1]{\ensuremath{\mathsf{#1}}}

\newcommand{\avg}[1]{\left< #1 \right>}
\newcommand{\ie}{i.e.}

 \biboptions{sort&compress}

\journal{Chemical Engineering Science}

\begin{document}

\begin{frontmatter}



\title{Brownian dynamics simulations of planar mixed flows of polymer solutions at finite concentrations}

\author[label1]{Aashish Jain \fnref{fn1}}
\author[label1]{Chandi Sasmal}
\author[label2]{Remco Hartkamp \fnref{fn2}}
\author[label4]{B. D. Todd}
\author[label1]{J.~Ravi~Prakash}
\address[label1]{Department of Chemical Engineering, Monash University, Clayton, VIC 3800, Australia}
\address[label2]{Multi Scale Mechanics, MESA+ Institute for Nanotechnology, University of Twente, P.O. Box 217, 7500 AE Enschede, The Netherlands}
\address[label4]{Department of Mathematics, Faculty of Science, Engineering and Technology, Swinburne University of Technology, Hawthorn, VIC 3122, Australia}
\fntext[fn1]{Current address: Department of Chemical Engineering and Materials Science, University of Minnesota,  Minneapolis, Minnesota 55455, USA}
\fntext[fn2]{Current address: MultiScale Material Science for Energy and Environment, CNRS/MIT (UMI 3466), Department of Civil and Environmental Engineering, Massachusetts Institute of Technology, Cambridge, MA 02139, USA}

\begin{abstract}
Periodic boundary conditions for planar mixed flows are implemented in the context of a multi-chain Brownian dynamics simulation algorithm. The effect of shear rate $\dot{\gamma}$, and extension rate $\dot{\epsilon}$, on the size of polymer chains, $\left<R_e^2\right>$, and on the polymer contribution to viscosity, $\eta$, is examined for solutions of FENE dumbbells at finite concentrations, with excluded volume interactions between the beads taken into account. The influence of the mixedness parameter, $\chi$, and flow strength, $\dot{\Gamma}$, on $\left<R_e^2\right>$ and  $\eta$, is also examined, where $\chi \rightarrow 0$ corresponds to pure shear flow, and $\chi \rightarrow 1$ corresponds to pure extensional flow. It is shown that there exists a critical value, $\chi_\text{c}$, such that the flow is shear dominated for $\chi < \chi_\text{c}$, and extension dominated for $\chi > \chi_\text{c}$. 
\end{abstract}
\begin{keyword}
polymer solutions \sep planar mixed flows \sep polymer contribution to viscosity \sep Brownian dynamics simulations 


\end{keyword}

\end{frontmatter}


\section{Introduction}
\label{Sec:Intro}
The study of the rheological behaviour of polymer solutions under different flow conditions has always been of great interest to the rheology community, both from a fundamental, and a practical point of view~\cite{birdetal1,LarsonBook1999}. The most commonly studied flows are shear and elongational flows because of their simplicity. They have proven to be useful in understanding many industrial processes such as extrusion, injection molding and sheet casting, to name but a few~\cite{baird1998polymer}. In many practical situations, however, rather than only shear or elongational flow, a combination of these flows is often observed. A special case is the linear combination of shear and elongational flow, the so-called \emph{mixed flow}~\cite{FullerLealJNNFM1981,HurPRE2002,WooShaqfehJCP2003,DuaJCP2003,HoffmanShaqfehJoR2007}. While elongational flows are shear free flows, shear flows have equal contributions from vorticity and elongation. In mixed flows both elongational and rotational components exist but their contributions vary, characterized by a \emph{mixedness} parameter $\chi$. In the limit $\chi \rightarrow 0$, the flow reduces to shear flow, while the limits $\chi \rightarrow -1$ and $\chi \rightarrow 1$, correspond to pure rotational and pure elongational flow, respectively. Experimentally, mixed flows have been generated and studied using the four-roll mill \citep{MullerAPL2007}. While there have been relatively few computational studies of mixed flows of dilute polymer solutions \citep{HurPRE2002,WooShaqfehJCP2003,DuaJCP2003,HoffmanShaqfehJoR2007}, there have been almost no computational studies of polymer solutions at finite concentrations undergoing mixed flow. Such flows are of significant interest in many practical applications, particularly in situations where there is a strong elongational component to the deformation, such as in inkjet printing or fibre spinning~\cite{xu07,zettl09}. Consequently, obtaining a quantitative understanding of the rheological behaviour of  non-dilute polymer solutions is not only of fundamental importance, but also vitally important for a number of practical applications. The aim of this paper is to develop a computational algorithm that enables the simulation of polymer solutions at finite concentrations subjected to planar mixed flows.

A challenging aspect of the development of an algorithm to simulate flows of finite-concentration polymer solutions, is the implementation of appropriate periodic boundary conditions (PBCs), arising from the need to carry out simulations for an indefinitely long time. PBCs for planar shear flows and planar elongational flows have been developed by \citet{LeesEdwardsJPC1972} and \citet{KraynikReineltIJMF1992}, respectively, that enable computations to run indefinitely in these flows. These PBCs have, for example, been used by \citet{BhupathirajuMPhys1996} and \citet{ToddDaivisPRL1998} in nonequilibrium molecular dynamics (NEMD) simulations. Apart from NEMD simulations, these PBCs have also been implemented in a Brownian dynamics (BD) simulation algorithm by \citet{StoltJoR2006} to simulate semidilute polymer solutions undergoing planar shear and planar elongational flows. In the context of planar mixed flows, \citet{WooShaqfehJCP2003, DuaJCP2003} and \citet{HoffmanShaqfehJoR2007} have carried out simulations of dilute polymer solutions using a BD algorithm. However, PBCs are \emph{not} required in single chain simulations. \citet{HuntJCP2010} have derived suitable PBCs for planar mixed flows and implemented them in an NEMD algorithm, which has recently been applied in a couple of different contexts~\citep{HartkampJCP2012,HartkampJCP2013}  

While NEMD simulations have led to important insights into the behaviour of polymer melts in a variety of flows~\cite{todd2001computer,kroger2004simple,hajizadeh2014} they are not suited to simulating the large-scale and long-time behaviour of solutions of  long polymer chains, because of the large number of degrees of freedom involved, and because such systems typically have relaxation times that are of the order of several seconds. Basically, the need to resolve the uninteresting motions of all the solvent molecules for extended periods of time, makes NEMD simulations computationally expensive and inefficient. It is generally accepted that the best approach under these circumstances is to use mesoscopic simulation algorithms, such as the hybrid LB/MD~\cite{BurkhardJCP1999}, or MPCD~\cite{GompperAPS2009} algorithms, or Brownian dynamics, in which the solvent molecules are discarded altogether and treated implicitly. 

To our knowledge, mixed flow PBCs have not been implemented in the context of a BD algorithm so far. In this paper, we discuss the implementation of PBCs for planar mixed flows in a multi-chain BD algorithm.  In particular, we adapt the PBC implementation in NEMD by \citet{HuntJCP2010} to the context of BD. The development of such an algorithm will enable the simulation of the large-scale and long-time properties of polymer solutions in industrially relevant flows at industrially relevant concentrations. 

To illustrate the capabilities of the BD algorithm developed here, we present some preliminary results on the planar mixed flow of non-dilute polymer solutions. Shaqfeh and coworkers \citep{HurPRE2002,WooShaqfehJCP2003,HoffmanShaqfehJoR2007}, have shown that the mixedness parameter $\chi$ is essential to understanding the nature of polymer behaviour in mixed flows. For instance, $\chi$ is a key parameter in determining the existence of the phenomenon of coil-stretch hysteresis~\cite{de-GennesJCP1974,SchroederScience2003,SchroederMacro2004}. Here, we study the influence of flow type $\chi$, and flow strength $\dot \Gamma$ on the viscosity in planar mixed flows, using the definition of viscosity introduced by \citet{HounkonnouJCP1992}. Additionally we show that, as in the case of dilute solutions, there exits a critical value, $\chi_c$, below which the flow is shear dominated, while being extension dominated for $\chi > \chi_c$. We find that the concentration of the polymer solution influences $\chi_c$, and consequently the nature of the flow.

The plan of the paper is as follows. Different forms of the velocity gradient tensor for planar mixed flows are discussed in section \ref{Sec:PMF}. In section \ref{Sec:SimAlgo} we discuss the governing equations of the BD algorithm (section \ref{SubSec:BDS}), the implementation of PBCs in planar mixed flows (section \ref{SubSec:PBC}), the definition of various macroscopic properties (section \ref{SubSec:MacroProps}), and the validation of the BD algorithm by comparison with known results (section \ref{SubSec:Validation}). In section \ref{Sec:Results}, the results of simulations of FENE dumbbells are presented, and the influence of flow strength and mixedness parameter on polymer size and viscosity is discussed. The central conclusions of this work are summarised in section \ref{Sec:Conc}.

\section{Planar mixed flows}
\label{Sec:PMF}
The velocity gradient tensor for planar shear flow (PSF) in matrix form is~\cite{birdetal1},
\begin{equation}
\label{eq:PSFGrad}
{(\bm{\nabla v})}_{\text{PSF}} =
\left( {\begin{array}{ccc}
 0 & 0 & 0 \\
 \dot{\gamma} & 0 & 0  \\
 0 & 0 & 0  \\
 \end{array} } \right)
\end{equation}
where, $\dot{\gamma}$ is the shear rate. The simplicity of planar shear flows has motivated  many studies that have compared experimental observations with simulation predictions \citep{LarsonBook1999,LarsonJoR2000,LarsonJoR2004,SchroederMacro2005}.

The velocity gradient tensor for planar elongational flow (PEF) is given by~\cite{birdetal1},
\begin{equation}
\label{eq:PEFGrad}
{(\bm{\nabla v})}_{\text{PEF}} =
\left( {\begin{array}{ccc}
 \dot{\epsilon} & 0 & 0 \\
 0 & -\dot{\epsilon} & 0  \\
 0 & 0 & 0  \\
 \end{array} } \right)
\end{equation}
where $\dot{\epsilon}$, is the elongational rate. Planar elongational flows occur in many industrial processes, and are generally difficult to study using computer simulations and experimental techniques, since in PEF, fluid elements are stretched exponentially with time in one direction while being contracted in the perpendicular direction~\cite{birdetal1}, leading to a very short span of time in which to observe the phenomena of stretching. 

In planar mixed flow (PMF), the velocity gradient tensor has the following form~\cite{FullerLealJNNFM1981,HounkonnouJCP1992,HoffmanShaqfehJoR2007,HuntJCP2010}
\begin{equation}
\label{eq:PMFGradGE}
{(\bm{\nabla v})}_{\text{PMF}} =
\left( {\begin{array}{ccc}
 \dot{\epsilon} & 0 & 0 \\
 \dot{\gamma} & -\dot{\epsilon} & 0  \\
 0 & 0 & 0  \\
 \end{array} } \right)
\end{equation}
which is referred to as the \emph{canonical form} \citep{HuntJCP2010}. The expanding direction is along the $x$-axis and the contracting direction is along the $y$-axis, with elongational field strength $\dot{\epsilon}$, while the shear gradient is along the $y$ direction, with shear field strength $\dot{\gamma}$. It follows that the expansion axis is always parallel to the $x$-axis, but the contraction axis is along the direction of one of the eigenvectors of the velocity gradient tensor. While the form of the velocity gradient tensor given by ${(\bm{\nabla v})}_{\text{PMF}}$ [Eq. (\ref{eq:PMFGradGE})] instinctively separates the shear and elongational flow components, it does not permit one to easily study the variation in material behavior as the flow changes smoothly from pure shear to pure elongation or vice versa. 

An alternative version of the velocity gradient tensor ${(\bm{\nabla v})}$ proposed by \citet{FullerLealJNNFM1981},
\begin{equation}
\label{eq:VelGradFuller}
{(\bm{\nabla v})} =
\left( {\begin{array}{ccc}
 0 & \dot{\Gamma} \chi & 0 \\
 \dot{\Gamma} & 0 & 0  \\
 0 & 0 & 0  \\
 \end{array} } \right)
\end{equation}
where $\dot{\Gamma}$ is the characteristic strain rate, and $\chi$ ($\in [-1,1]$) is the mixedness parameter (which measures the relative strength of rotational and elongational components), is more suited to this purpose. It can be shown that this form for ${(\bm{\nabla v})}$ reduces to PSF when $\chi \rightarrow 0$, while  pure PEF is recovered in the limit $\chi \rightarrow 1$. Eq. (\ref{eq:VelGradFuller}) is also valid in the limit of $\chi \rightarrow -1$, which corresponds to the pure rotational flow limit. 

In their studies of PMF of dilute polymer solutions, \citet{HoffmanShaqfehJoR2007} have shown that Eq. (\ref{eq:VelGradFuller}) is equivalent to 
\begin{equation}
\label{eq:VelGradShaq}
{(\bm{\nabla v})} =
\left( {\begin{array}{ccc}
 \dot{\Gamma} \sqrt{\chi} & 0 & 0 \\
 \dot{\Gamma} (1 - \chi) & -\dot{\Gamma} \sqrt{\chi} & 0  \\
 0 & 0 & 0  \\
 \end{array} } \right)
\end{equation}
in a suitably rotated coordinate system, where they confine their attention to elongation-dominated mixed flow, for which $\chi > 0$. Clearly, ${(\bm{\nabla v})}$ and ${(\bm{\nabla v})}_{\text{PMF}}$ are similar in structure. Comparing Eqs. (\ref{eq:PMFGradGE}) and (\ref{eq:VelGradShaq}), we can express the shear rate $\dot{\gamma}$ and elongational rate $\dot{\epsilon}$ in terms of $\dot{\Gamma}$ and $\chi$ as follows
\begin{equation}
\label{eq:gdot-chi}
\dot{\gamma} = \dot{\Gamma} (1 - \chi)
\end{equation}
and
\begin{equation}
\label{eq:edot-chi}
\dot{\epsilon} = \dot{\Gamma} \sqrt{\chi}
\end{equation}
The smooth crossover between pure planar shear and pure planar elongational flow limits can be studied by varying $\chi$ between $0$ and $1$. 
\section{Polymer model and simulation algorithm}
\label{Sec:SimAlgo}

A linear bead-spring chain model \citep{BirdVol2} is used to represent polymers at the mesoscopic level, with each polymer chain coarse-grained into a sequence of $N_b$ beads, which act as centers of hydrodynamic resistance, connected by $N_b - 1$ massless springs that represent the entropic force between adjacent beads. A finite-concentration polymer solution is modeled as an ensemble of such bead-spring chains, immersed in an incompressible Newtonian solvent. A total of $N_c$ chains are initially enclosed in a cubic and periodic cell of edge length $L$, giving a total of $N = N_b \times N_c$ beads per cell at a bulk monomer concentration of $c = N/V$, where $V = L^3$ is the volume of the simulation cell.
\subsection{BD simulations of flowing  polymer solutions at finite concentration}
\label{SubSec:BDS}
The Euler integration algorithm, in the absence of hydrodynamic interactions, for the non-dimensional Ito stochastic differential equation governing the position vector ${\Vector{r}}_{\nu}(t)$ of bead $\nu$ at time $t$, is~\cite{StoltJoR2006,JainPRE2012},
\begin{eqnarray}
\label{eq:sde}
{\Vector{r}}_{\nu}(t+\Delta t)  &=& 
{\bm{\Vector{r}}}_{\nu}(t) + \bm{\Tensor{\kappa}} 
\cdot \bm{\Vector{r}}_{\nu}(t) 
+ \left( \frac{1}{4} \right) {\bm{\Vector{F}}}_{\nu}(t) \, \Delta t \nonumber \\ 
&+& \left( \frac{1}{\sqrt{2}} \right) {\Vector{\Delta \bm{W}}}_{\nu}(t)
\end{eqnarray}
The length scale $l_H = \sqrt{k_B T/H}$ and time scale ${\lambda}_H = \zeta/4H$ (where, $k_B$ is the Boltzmann constant, $T$ is the temperature, $H$ is the spring constant and
$\zeta$ is the hydrodynamic friction coefficient associated with a bead), have been used for the purpose of non-dimensionalising Eq~(\ref{eq:sde}). The $3 \times 3$ tensor $\bm{\Tensor{\kappa}}$ is equal to ${(\bm{\nabla v})}^T$, with $\bm{v}$ being the unperturbed solvent velocity. ${\bm{\Vector{F}}}_{\nu}$ incorporates all the non-hydrodynamic forces on bead $\nu$ due to all the other beads. The non-hydrodynamic forces in the model are comprised of the spring forces ${\bm{\Vector{F}}}_{\nu}^{\text{spr}}$ and excluded volume interaction forces ${\bm{\Vector{F}}}_{\nu}^{\text{exv}}$, ie.,
${\bm{\Vector{F}}}_{\nu} = {\bm{\Vector{F}}}_{\nu}^{\text{spr}} + {\bm{\Vector{F}}}_{\nu}^{\text{exv}}$. 
The components of the Gaussian noise $\Vector{\Delta  \bm{W}}_{\nu}$ are obtained from a real-valued Gaussian distribution with zero mean and variance $\Delta t$. 

The specification of the force term in Eq.~(\ref{eq:sde}) requires the consideration of bonded and non-bonded interactions between beads, with the former arising due to the presence of spring forces. In order to model spring forces, a finitely extensible nonlinear elastic (FENE) potential has been used. The entropic spring force on bead $\nu$ due to adjacent beads can be expressed as
${\bm{\Vector{F}}}_{\nu}^{\text{spr}} = 
{\bm{\Vector{F}}}^c({\bm{\Vector{Q}}}_{\nu}) -
{\bm{\Vector{F}}}^c({\bm{\Vector{Q}}}_{\nu - 1})$ where
${\bm{\Vector{F}}}^c({\bm{\Vector{Q}}}_{\nu - 1})$ is the force between the beads $\nu -1$ and $\nu$, acting in the direction of the connector vector between the two beads ${\bm{\Vector{Q}}}_{\nu - 1} =
{\bm{\Vector{r}}}_{\nu} - {\bm{\Vector{r}}}_{\nu - 1}$. 
The dimensionless FENE spring force is given by ${\bm{\Vector{F}}}^c({\bm{\Vector{Q}}}_{\nu}) = \dfrac{{\bm{\Vector{Q}}}_{\nu}}{1-{\vert\bm{\Vector{Q}}}_{\nu}\vert^2/b}$, where $b = H q_0^2 /k_B T$ is the dimensionless finite extensibility parameter, and $q_0$ is the dimensional maximum stretch of a spring. 

In this paper we consider only excluded volume interactions as the source of non-bonded interactions.  The excluded volume interactions are modeled using a narrow Gaussian potential \citep{RaviOttingerMacro1999,ottinger,prakash:macromol-01}, which in terms of non-dimensional variables is given by
\begin{equation}
E \left( \Vector{r}_{\nu \mu} \right) = z^{\star}  \, \left(   \frac{1}{ {d^{\star}}^3} \right)  
\exp \left \lbrace - \frac{1 }{2}\, \frac{ \Vector{r}_{\nu \mu}^2 }{ {d^{\star}}^2} \right \rbrace
\label{evpot1}
\end{equation}
The dimensionless parameter $z^\star$ is the strength of excluded volume interactions and $d^\star$ is a dimensionless parameter that measures the range of the excluded volume interaction. $z^\star$ is related to the solvent quality parameter $z$ through $z^\star = z/\sqrt{N_b}$, and $d^\star$ is related to $z^\star$ through $d^\star = K {z^\star}^{1/5}$, where $K$ is an arbitrary parameter which becomes irrelevant in the long chain limit \citep{RaviCES2001,KumarPrakashMacro2003}. We have used a value of $K=1$ in all the simulation results reported here. 

While the implementation of the term $\left[\bm{\Tensor{\kappa}} \cdot \bm{\Vector{r}}_{\nu}(t) \right]$ in Eq.~(\ref{eq:sde}) is straightforward, the major challenge is in the implementation of appropriate periodic boundary conditions for various flows. Periodic boundary conditions (PBCs)  are used in simulations to mimic real systems, enabling the computation of bulk properties by simulating only a small number of particles. The implementation of PBCs for planar mixed flows in the context of BD simulations is discussed in the section below.

In order to compare BD simulation predictions with experimental observations on polymer solutions, it is essential to include hydrodynamic interactions (HI) in the simulation algorithm~\cite{RaviBookchapter1999,prabhakar:jor-04,SuntharMacro2005,SuntharEPL2006,RaviKARJ2009,JainPRL2012}. We have previously discussed the development of optimised BD algorithms with HI in the context of both dilute and semidilute solutions~\cite{PrabhakarJNNFM2004,JainPRE2012}. In the present instance, we have neglected HI since we want to focus on the aspects dealing with the implementation of mixed flow PBCs. It turns out it is sufficient to include pair-wise non-linear excluded-volume interactions in order to invoke all the aspects of the algorithm that are related to the implementation of PBCs in flow. This is discussed in greater detail in section~\ref{SubSec:Validation} where we consider the validation of the current BD algorithm. 

\subsection{Periodic boundary conditions for planar mixed flows}
\label{SubSec:PBC}
In flow simulations, PBCs require that the shape of the simulation box changes with time in accordance with the flow such that the deformation of the simulation box follows the streamlines of the flow. As the simulation box deforms with respect to time, there comes a time when the box has deformed to such an extent that the minimum spacing between any two sides of the box becomes less than twice the inter-particle interaction range. At that point in time, particles start to interact with themselves and the simulation needs to be stopped. There might also be cases, such as in shear flows, where after some time, one of the sides of the box becomes very large resulting in numerical problems. In other words, the deformation of the simulation box in such a manner restricts the simulation from proceeding for long times. In fact, this issue becomes even more serious for polymer molecules, since in this case, relaxation times in general are quite long, and it is very important to simulate them for sufficiently long time in order to capture their dynamics accurately. It is consequently necessary to perform a mapping of the simulation box such that the initial box configuration is periodically recovered. Remapping of the box configuration requires two conditions to be met: (i) \emph{Compatibility}, which means that the minimum lattice spacing should never be less than twice the range of inter-particle interactions, and (ii) \emph{Reproducibility}, which means that the lattice points of a lattice should overlap with the lattice points of the same lattice at a different time. Remapping of the lattice in NEMD simulations of planar shear flow was first carried out by \citet{EvansMolPhys1979} and \citet{EvansMolSim1994} who modified the original sliding-brick algorithm of \citet{LeesEdwardsJPC1972} to a deforming-box algorithm. Satisfying the two conditions of compatibility and reproducibility, \citet{KraynikReineltIJMF1992} developed PBCs capable of being remapped, for planar elongational flows. The Kraynik-Reinelt PBCs were first implemented by \citet{ToddDaivisPRL1998} and \citet{BaranyaiCummingsJCP1999} in their planar elongational NEMD simulation algorithms. In these PBCs, basically the lattice is started at an angle $\theta$ (the so-called magic angle) \citep{KraynikReineltIJMF1992}, then deformed for a certain period of time $\tau_p$ (the strain period), and then mapped back to its original state. This process of deforming the lattice till $\tau_p$ and mapping back to its original state is repeated as many times as needed, to achieve extended simulations. 

\citet{HuntJCP2010} extended the PBCs for planar elongational flows to planar mixed flows in their NEMD simulations for the first time. In this paper, we adopt the reproducible periodic boundary conditions for planar mixed flow developed by \citet{HuntJCP2010}, and use it in a multi-chain Brownian dynamics simulation algorithm for semidilute polymer solutions. Implementation of PBCs for planar mixed flow is similar to that for planar elongational flow \citep{KraynikReineltIJMF1992,ToddDaivisPRL1998}, except for some differences due to the presence of a rotational component. These differences are briefly outlined below, along with the major steps in the implementation of PBCs for PMF.

\begin{figure}[!tbp]
\centering
\resizebox{7cm}{!} {\includegraphics{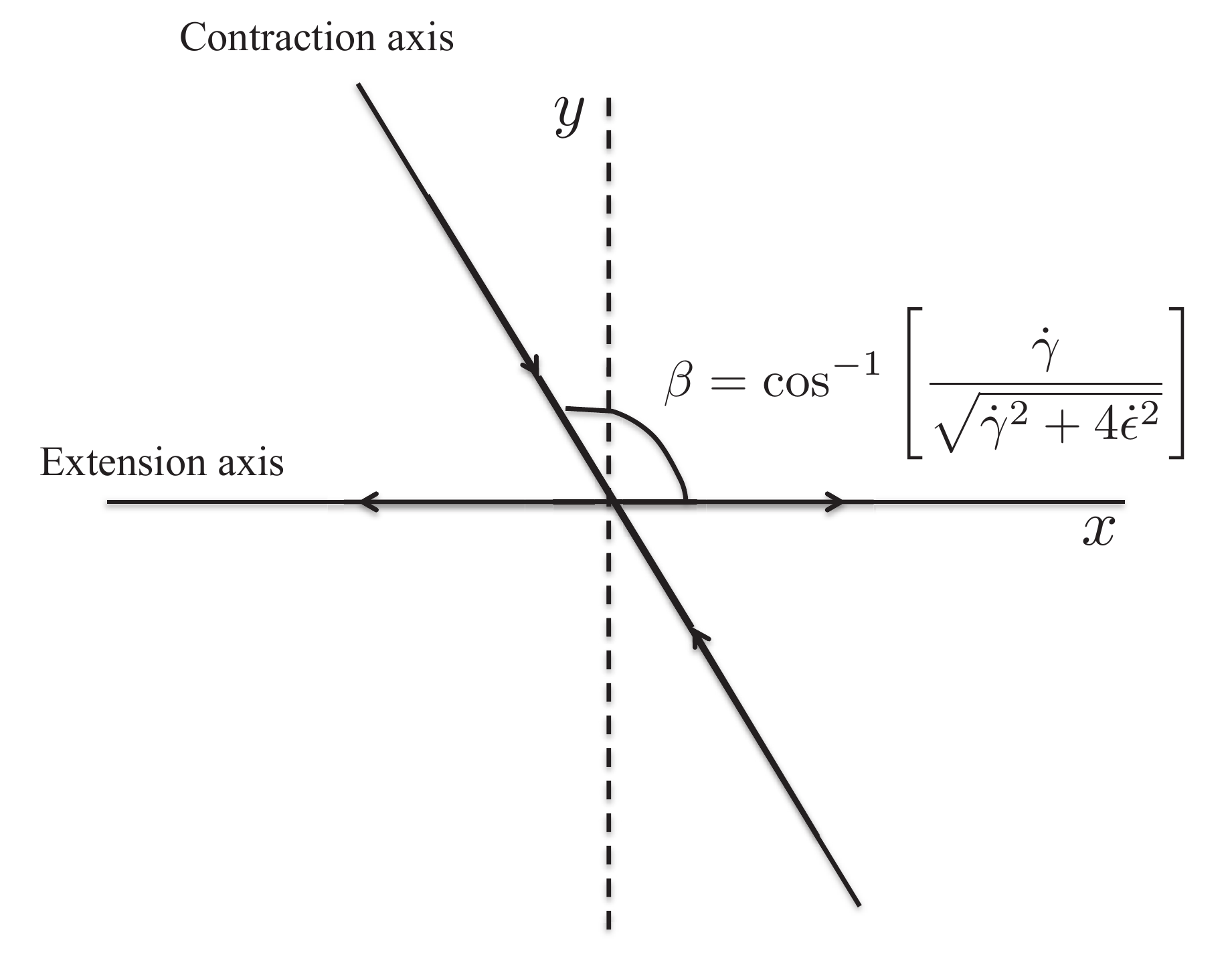}}
\caption{Extension and contraction axes in planar mixed flow.}
\label{MixedFlowAxis}
\end{figure}

In the canonical representation, the eigenvalues of ${(\bm{\nabla v})}_{\text{PMF}}$ are $\{\dot{\epsilon}, -\dot{\epsilon}, 0\}$, and a possible choice of the corresponding eigenvectors is ($1$, $\dot{\gamma}/2 \dot{\epsilon}$, $0$), $\left(0, 1, 0\right)$ and $\left(0, 0, 1\right)$.
It is worth noting that the eigenvalues of the velocity gradient tensor of the canonical PMF are equivalent to those for PEF, where ${(\bm{\nabla v})}_{\text{PEF}}$ is already in a diagonal form. However, the eigenvectors corresponding to the eigenvalue $\dot{\epsilon}$ are different for ${(\bm{\nabla v})}_{\text{PMF}}$ and ${(\bm{\nabla v})}_{\text{PEF}}$. For PEF, the eigenvector corresponding to $\dot{\epsilon}$ is ($1$, $0$, $0$), which leads to the fact that the extension axis and contraction axis are orthogonal. In case of the canonical PMF, the eigenvector corresponding to $\dot{\epsilon}$ is ($1$, $\dot{\gamma}/2 \dot{\epsilon}$, $0$), resulting in a system where the extension axis and contraction axis are non-orthogonal (for simplicity, we will henceforth refer to ``canonical" PMF as PMF). In PMF, the angle $\beta$ between the extension axis and the contraction axis (displayed in Fig.~\ref{MixedFlowAxis}), depends on the ratio of $\dot{\gamma}$ to $\dot{\epsilon}$ since, \begin{equation}
\label{eq:betaPMF}
\beta = {\cos}^{-1} \left[\frac{\dot{\gamma}}{\sqrt{{\dot{\gamma}}^2 + 4 {\dot{\epsilon}}^2}}\right]
\end{equation}

Two important parameters with regard to PBC implementation in flow are the magic angle and strain period, both of which depend on the eigenvalues of the velocity gradient tensor. Since the eigenvalues of the velocity gradient tensor for PMF and PEF are the same, the magic angle and strain period for PMF can be obtained in a similar manner as in the case of PEF. However, the initial lattice configuration for PMF is different from that of PEF because of the differences in the eigenvectors discussed above. Details of the initial lattice vectors for both PEF and PMF can be found in Ref.~\citenum{JainThesis}. 

For the sake of completeness, the derivation of initial lattice vector for PMF is discussed in~\ref{Sec:Ap}. Here, the steps involved in the implementation of PBCs for PMF are briefly enumerated below:
\begin{enumerate}
\item An integer value of the parameter $\tilde k$ (which controls both the magic angle and the strain period) is chosen, such that $\tilde k \ge 3$, and $\tilde k \in [3, 4, 5, \ldots]$.
\item The eigenvalue $\phi$ (as defined in Ref.~\citenum{KraynikReineltIJMF1992}) is calculated using the expression,
\begin{equation}
\phi = \frac{\tilde k + \sqrt{\tilde k^2 - 4}}{2} 
\end{equation}
\item The strain period $\tau_p$ is estimated using the expression $\tau_p = \log{(\phi)}/\dot{\epsilon}$, where $\dot{\epsilon}$ is the elongational rate.
\item A choice is made for the values of $N_{11}$ and $N_{12}$, which are the ``11'' and ``12'' elements of a $3 \times 3$ integer  matrix that describes the mapping between the deformed and original matrix, as follows. Basically, a positive integer value of $N_{11}$ is selected such that an integer value of $N_{12}$ is obtained using the expression,
\begin{equation}
N_{12} = -\sqrt{N_{11} (\tilde k - N_{11}) - 1} 
\end{equation} 
Various possible values of  $N_{11}$ and $N_{12}$ are listed in Ref.~\citenum{KraynikReineltIJMF1992}.
\item Finally, the magic angle is calculated from,
\begin{equation}
\theta = {\tan}^{-1} \left[\frac{N_{11} - \phi}{N_{12}}\right] 
\end{equation}
\end{enumerate}
Using initial lattice vectors that depend on the magic angle, the simulation can be started and run until the strain period. 
The lattice is then mapped back to its original state, and this way the simulation can be carried out for an extended period. 

With regard to the compatibility condition, as discussed earlier, there is an issue with the length of one of the sides of the simulation box decreasing with time. \citet{KraynikReineltIJMF1992} have shown that the reproducibility condition automatically guarantees the compatibility condition, \ie, they have shown that the distance $D(\tau_p)$ between any two lattice points at time $\tau_p$ is never less than the minimum lattice spacing $D_{\text{min}}$, such that the lattice points do not overlap. 
In simulations, the cutoff radius of any inter-particle interaction potential is always chosen to be less than $D_{\text{min}}/2$, which ensures that the compatibility condition is always satisfied. The derivation of $D_{\text{min}}$ for PMF has been discussed by~\citet{HuntJCP2010}.

\subsection{Macroscopic properties}
\label{SubSec:MacroProps}
Static and dynamic properties of polymer solutions at equilibrium can be calculated once the trajectories of the time evolution of all the beads on all the chains are obtained using Eq.~(\ref{eq:sde}). For rheological properties, not only are the bead configurations required, it is also necessary to know the forces acting on them.

Two important equilibrium static properties are (i) the end-to-end distance, and (ii) the gyration radius, which are used to assess the mean dimension of a polymer chain \citep{DoiEdwardsBook,RubinsteinBook}. The end-to-end distance is defined as the mean square distance between the first and the last beads on a chain,
\begin{equation}
\label{eq:re2single}
\avg{R_{e}^{2}} = \avg{(\Vector{r}_{N_b} - \Vector{r}_0)^2}
\end{equation}
where, $\langle \cdot \cdot \cdot \rangle$ represents an ensemble average, and $\Vector{r}_0$ and $\Vector{r}_{N_b}$ are position vectors of the first and the last bead, respectively. 
The mean square gyration radius, which is the mean square distance between the beads and the center of mass $\Vector{r}_{\text{cm}}$ of the chain is defined by,
\begin{equation}
\label{eq:rg2single}
\avg{R_{g}^{2}} =  \frac{1}{N_b} \sum_{\mu=1}^{N_b} \avg{(\Vector{r}_{\mu} - \Vector{r}_{\text{cm}})^2}
\end{equation}
where, $\Vector{r}_{\text{cm}} = \dfrac{1}{N_b} \sum_{\mu=1}^{N_b} \Vector{r}_{\mu}$.

The behaviour of polymer solutions, when subjected to an imposed flow, is described by various material functions that are defined in terms of the components of the stress tensor~\cite{birdetal1}. In the absence of external forces, the stress tensor (non-dimensionalized by $n_p \,k_B T$, where $n_p$ is number of polymer chains per unit volume), for a multi-chain system can be shown to be~\cite{StoltzThesis}, 
\begin{equation}
\label{eq:taup1}
\bm{\sigma} = \frac{1}{N_c} \, \left[ \sum_{\mu = 1}^{N} \sum_{\nu = 1}^{N}  \avg{\bm{\Vector{r}}_{\mu \nu} {\bm{\Vector{F}}}_{\mu \nu}^{\text{exv}}} \, + \sideset{}{_{N_c}}\sum \sum_{\nu=1}^{N_b - 1} \avg{\Vector{Q}_{\nu} \Vector{F}^c (\Vector{Q}_{\nu}) }    \right]
\end{equation}
In the above equation, the first term is the contribution due to excluded volume forces among the beads, where $\bm{\Vector{r}}_{\mu \nu} = \bm{\Vector{r}}_{\mu}-\bm{\Vector{r}}_{\nu}$ is the vector between beads $\nu$ and $\mu$, and $\bm{\Vector{F}}_{\mu \nu}^{\text{exv}}$ is the excluded volume force between them. The second term is the contribution due to spring forces, where $\Vector{Q}_{\nu}$ is the connector vector between the two beads $\Vector{Q}_{\nu} = \Vector{r}_{\nu + 1} - \Vector{r}_{\nu}$, and $\Vector{F}^c (\Vector{Q}_{\nu})$ is the spring force between the beads $\nu$ and $\nu + 1$.

Once the stress tensor is calculated, the various solution material functions can be estimated. In this work, we have focused our attention on the polymer contribution to the solution's viscosity. \citet{HounkonnouJCP1992} have proposed the following expression for a generalized viscosity $\eta$ for any arbitrary flow gradient tensor, 
\begin{equation}
\label{eq:EtaH}
\eta  = \frac{\bm{\Tensor{\dot{\Gamma}}} : \Tensor{\bm{\sigma}}}{\bm{\Tensor{\dot{\Gamma}}} : \bm{\Tensor{\dot{\Gamma}}}} 
\end{equation}   
where $\bm{\Tensor{\dot{\Gamma}}}$ is the rate of strain tensor, defined by $\bm{\Tensor{\dot{\Gamma}}} = (\bm{\nabla v}) + (\bm{\nabla v})^T$. 
Using the definition of viscosity in~Eq. (\ref{eq:EtaH}), with $(\bm{\nabla v}) = {(\bm{\nabla v})}_{\text{PMF}}$ (see Eq.~(\ref{eq:PMFGradGE})), it is straightforward to show that in planar mixed flows the viscosity is given by,
\begin{equation}
\label{eq:EtaPMF}
\eta = -\frac{2 \dot{\epsilon} (\sigma_{xx} - \sigma_{yy}) + 2 \dot{\gamma} \sigma_{xy}}{8 \dot{\epsilon}^2 +  2 \dot{\gamma}^2} 
\end{equation}
In the limit of pure planar shear flow ($\dot{\epsilon} = 0$), Eq.~(\ref{eq:EtaPMF}) implies, 
\begin{equation}
\label{eq:etap-shear}
\eta_{\text{PSF}} = -\frac{\sigma_{xy}}{\dot{\gamma}}
\end{equation}
while in the limit of pure planar elongational flow ($\dot{\gamma} = 0$), Eq. (\ref{eq:EtaPMF}) leads to,
\begin{equation}
\label{eq:etap-elong}
\eta_{\text{PEF}} = -\frac{\sigma_{xx} - \sigma_{yy}}{4 \dot{\epsilon}}
\end{equation}
Note that this definition of the viscosity in planar extension flows differs from the conventional definition of the viscosity $\bar \eta_{1}$ used in the rheology literature~\cite{birdetal1}, 
\begin{equation}
\label{eq:etap-rheol}
\bar \eta_{1} = - \frac{\sigma_{xx} - \sigma_{yy}}{\dot{\epsilon}}
\end{equation}
since $\eta_{\text{PEF}} =  \bar \eta_{1}/4$. The advantage of the \citet{HounkonnouJCP1992} definition is that the generalized viscosity reduces to the Newtonian viscosity in the limit of either $\dot{\gamma} \to 0$, or $\dot{\epsilon} \to 0$. We use the \citet{HounkonnouJCP1992} definition in all our discussions of planar mixed flows. However, we use $\bar \eta_{1}$ when comparing results of the multi-chain algorithm with single chain simulations in planar extensional flows. 

From Eqs.~(\ref{eq:EtaPMF}) to~(\ref{eq:etap-elong}), the viscosity in planar mixed flows can be rewritten as a linear combination of $\eta_{\text{PSF}}$ and $\eta_{\text{PEF}}$,
\begin{equation}
\label{eq:etap-mixed}
\eta  = \frac{(4 \dot{\epsilon}^2 \eta_{\text{PEF}} + \dot{\gamma}^2 \eta_{\text{PSF}})}{4 \dot{\epsilon}^2 +   \dot{\gamma}^2} 
\end{equation}    
Eqs. (\ref{eq:etap-shear}) - (\ref{eq:etap-mixed}) have been used by \citet{HounkonnouJCP1992, BaranyaiJCP1995,ToddDaivisPRL1998,ToddJNNFM2003} and \citet{HuntJCP2010} in their NEMD simulations for the viscosity of various fluids. The PMF viscosity can also be expressed in terms of the strength of mixed flow $\dot{\Gamma}$, and the mixedness parameter $\chi$, by
\begin{equation}
\label{eq:Eta}
\eta  = -\frac{\sqrt{\chi} \, (\sigma_{xx} - \sigma_{yy}) + (1 - \chi) \, \sigma_{xy}}{\dot{\Gamma} \, \left[4 \,  {\chi} + (1 - \chi)^2 \, \right]} 
\end{equation}

In the current simulations of PMF, we use either Eq.~(\ref{eq:EtaPMF}) or  Eq.~(\ref{eq:Eta}) to calculate the viscosity, depending on whether we use the pair ($\dot \gamma, \dot \epsilon$), or ($\dot \Gamma, \chi$) to describe the flow.

\subsection{Validation of the BD algorithm}
\label{SubSec:Validation}

In order to validate the multi-chain BD flow algorithm, results are compared with the results from single-chain BD simulations in the dilute limit. Since the current algorithm is an extension of our previous equilibrium multi-chain BD algorithm \citep{JainPRE2012}, the new additional features in the flow algorithm are the implementation of (i) periodic boundary conditions for planar mixed flows, and (ii) a neighbour-list consistent with PBCs for PMF.

The neighbour-list and PBCs do not play a role in the flow simulation when hydrodynamic and excluded volume interactions are ignored. This situation corresponds to the Rouse model for which analytical expressions for various properties are known~\cite{BirdVol2}. 

As a simple test of the basic aspects of the algorithm (such as of the integrator with the flow term incorporated), hydrodynamic and excluded volume interactions are switched off, and the dimensionless mean square end-to-end distance $\langle R_e^2 \rangle$, of chains consisting of $10$ beads in the ultra dilute limit, is computed as a function of dimensionless shear rate $\dot{\gamma}$,  and compared  with the prediction of the Rouse model~\cite{BirdVol2},
\begin{equation}
\label{eq:rouseRe2}
{\langle R_e^2 \rangle}_{\text{Rouse}} = {\langle R_e^2 \rangle}_{\text{eq}} \, \left[1 + \frac{N_b \, (N_b + 1) \, (N_b^2 + 1) \, \dot{\gamma}^2}{45} \right] 
\end{equation}  
where, it may be recalled that ${\langle R_e^2 \rangle}_{\text{eq}} = 3 (N_b - 1)$ is the mean square end-to-end distance at equilibrium \citep{BirdVol2}. The factor of 3 in the expression for ${\langle R_e^2 \rangle}_{\text{eq}}$ comes from using a length scale $l_{H}$ which is $(1/3)^{\text{rd}}$ the mean square equilibrium size of a single spring. 

Note that the dilute and semidilute concentration regimes are demarcated by the overlap concentration $c^{\star}$, which is defined by, $c^{\star} = N_b / [({4\pi}/{3})\left(R^{0}_{g}\right)^3]$, where $R^0_g$ is the radius of gyration for an isolated chain at equilibrium. Results are reported in terms of the scaled variable, $c/c^\star$, which is calculated a priori by computing $R^0_g$ from single-chain BD simulations at equilibrium, for the relevant set of parameter values. In order to compare results of the multi-chain algorithm with dilute solution results, we typically choose extremely small values of $c/c^\star$ to prevent any likelihood of chain-chain interactions.

\begin{figure}[tbp]
\centering
\resizebox{7.5cm}{!} {\includegraphics{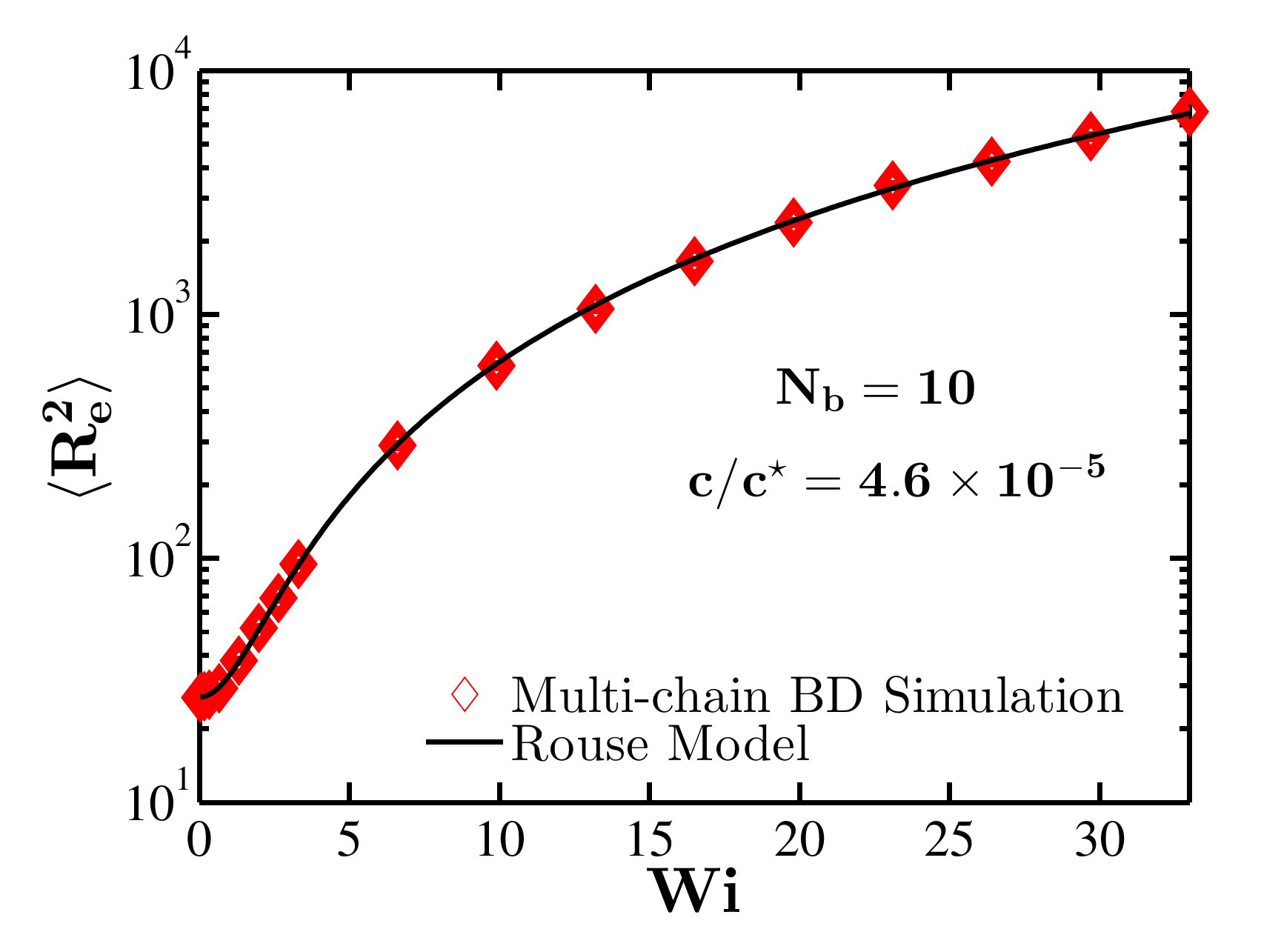}}
\caption{Mean square end-to-end distance obtained by BD simulations compared with Rouse model predictions, for bead-spring chains with $N_b = 10$ beads, as a function of the Weissenberg number $W\!i$.}
\label{RouseChainSizeVsShearRate}
\end{figure}

The red symbols in Fig.~\ref{RouseChainSizeVsShearRate} indicate the results for $\langle R_e^2 \rangle$ obtained by carrying out multi-chain BD simulations, while the solid line is the Rouse model prediction. The increase in the mean size of the chain with increasing strain rate, is represented in Fig.~\ref{RouseChainSizeVsShearRate} in terms of the Weissenberg number $W\!i = \lambda_{\eta} \dot{\gamma}$, which is a non-dimensional measure of the strain rate. The quantity $\lambda_{\eta}$ is a characteristic relaxation time defined by,
\begin{equation}
\label{eq:lambda}
\lambda_{\eta} = \frac{[\eta]_{0}M \eta_{s}}{N_{A}k_{B}T}
\end{equation}
where, $[\eta]_{0}$ is the zero shear rate intrinsic viscosity of the solution, $M$ is the molecular weight, $\eta_{s}$ is the solvent viscosity, and $N_{A}$ is Avagadro's number. One can show~\cite{RaviJoR2002}, in terms of the non-dimensionalisation scheme used here, that this definition implies that $\lambda_{\eta} = \eta_{0}$, where $\eta_{0}$ is the zero shear rate polymer contribution to the viscosity. Consequently, $W\!i = \eta_{0} \dot \gamma$. The Rouse model predicts a constant viscosity, independent of the shear rate, which can be calculated analytically to be~\cite{BirdVol2},
\begin{equation}
\label{eq:rouseeta}
\eta_\text{Rouse} = \frac{N_b^{2}-1 }{3} 
\end{equation}  
It follows that for a 10 bead chain, $\lambda_{\eta} = 33$. The multi-chain BD simulations were carried out with $N_c = 30$, $c/c^\star = 4.6 \times 10^{-5}$ and a time step size $\Delta t = 0.005$. Note that the Hookean spring force law, which corresponds to $b \rightarrow \infty$ in the FENE model, was considered in all the validation studies for planar shear flow. The excellent agreement between simulations and the Rouse model indicates that the algorithm is performing satisfactorily.    

\begin{figure}[tbp]
\begin{center}
\resizebox{7.5cm}{!} {\includegraphics*{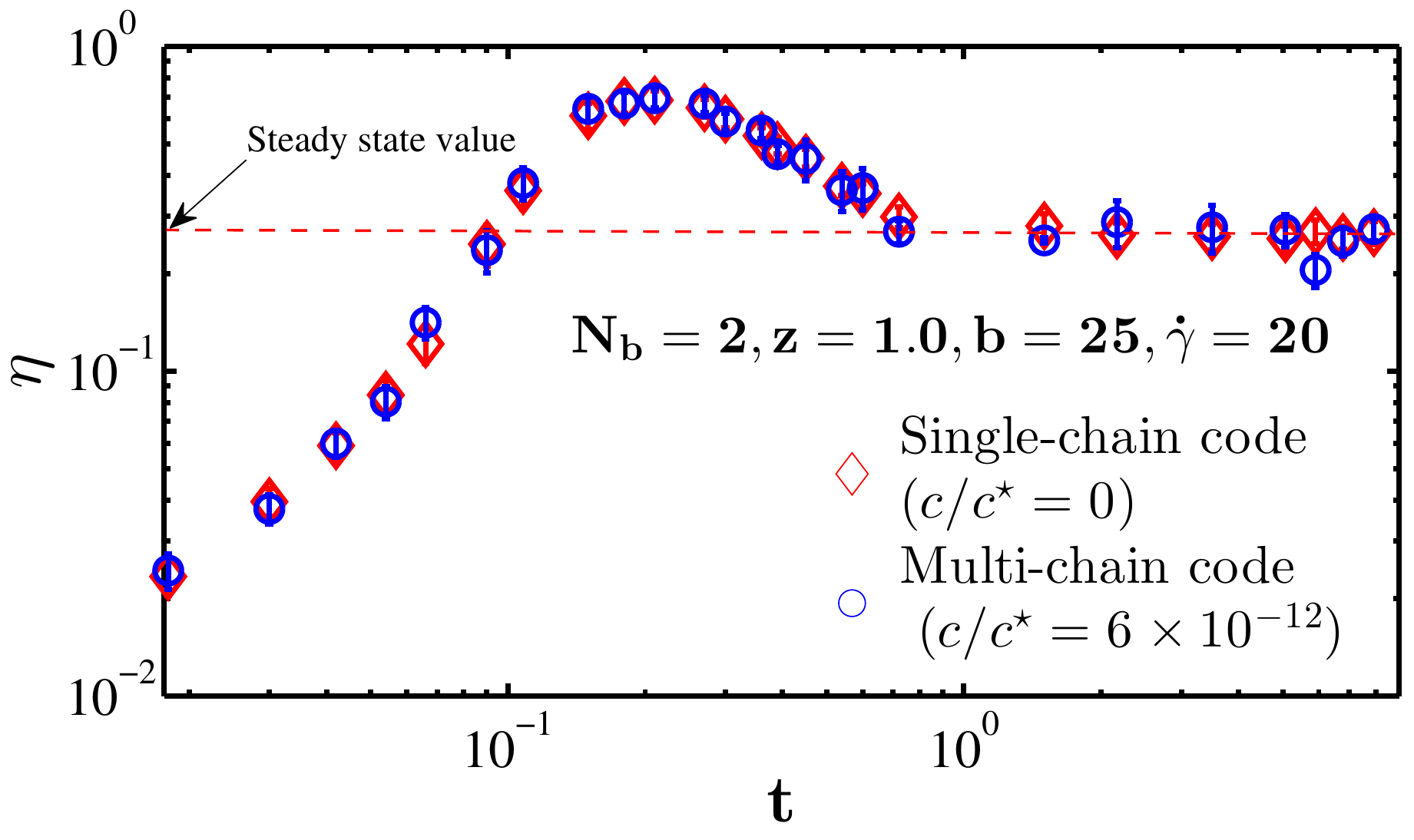}} 
\end{center}
\vskip-20pt
\caption{\small \label{transshear} Comparison of the transient viscosity upon inception of steady planar shear flow at non-dimensional times $t$, predicted by the multi-chain BD algorithm, with the results of single-chain BD simulations in the dilute limit. Excluded volume interactions are taken into account but hydrodynamic interactions are switched off. The parameter values are as indicated in the figure legend.}
\end{figure}

In order to test the neighbour-list and PBCs implementations, multi-chain BD simulations of dumbbells ($N_b = 2$) have been carried out in the ultra dilute limit, with excluded volume interactions between the dumbbell beads. In particular, we set $z = 1.7$, $N_c = 10$, $c/c^\star = 6 \times 10^{-12}$ and $\Delta t = 0.005$. A large number of independent runs (in the range of $10^3$ - $10^6$) were performed in order  to obtain results with acceptable error bars. We first examine the behaviour of the algorithm in transient flows, followed by steady state flows, in both planar shear and extensional modes of deformation. The former is important in order to ensure that there are no artifacts caused due to the periodic re-mapping of the system after every strain period. 
\begin{figure*}[tbp]
\centering
\begin{tabular}{cc}
\resizebox{8cm}{!}{\includegraphics{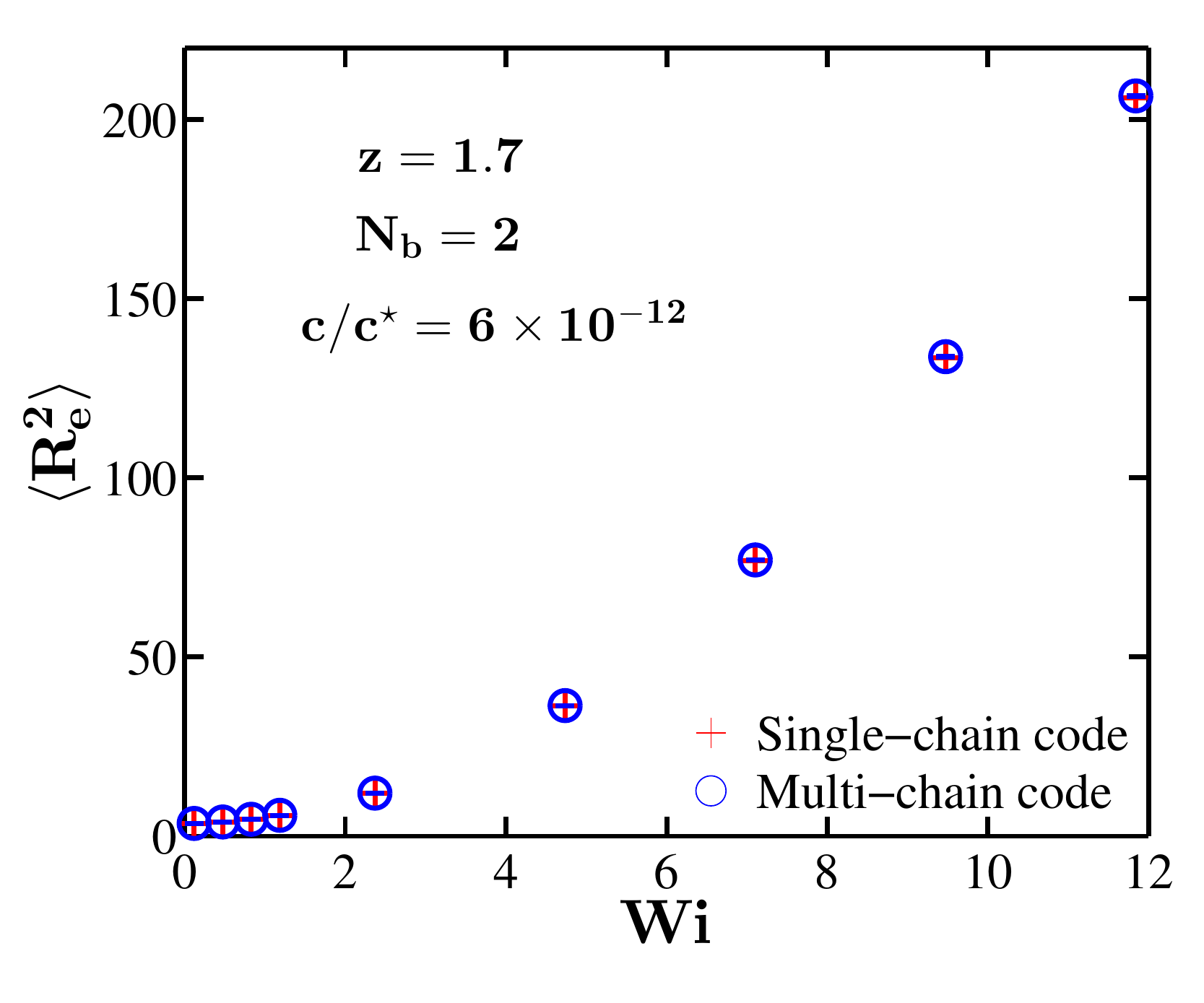}} &
\resizebox{8cm}{!}{\includegraphics{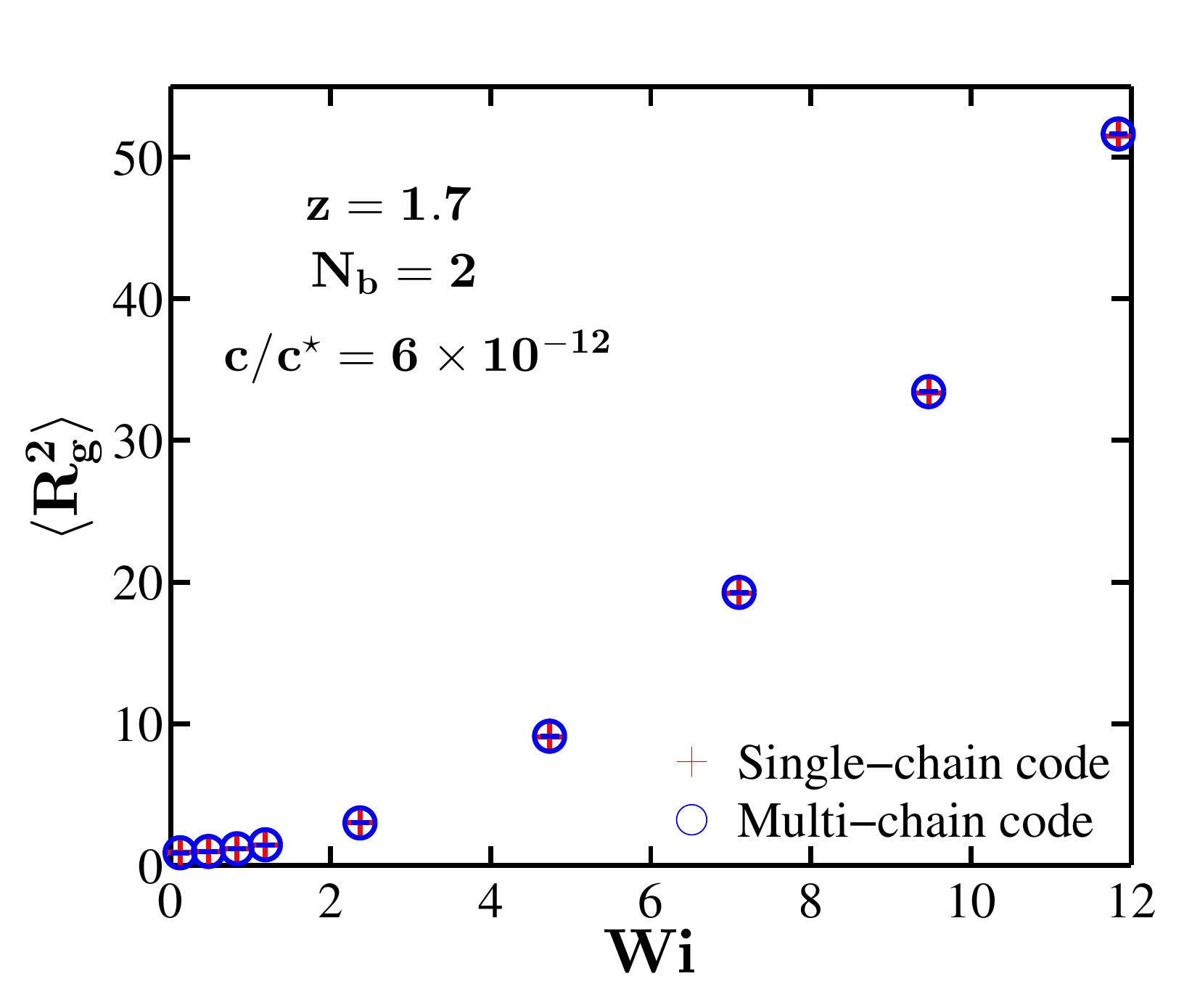}} \\
(a) & (b) \\  
\end{tabular}
\caption{\small \label{Re2Rg2PSFDilVal} Comparison of the mean square end-to-end distance $\langle R_e^2 \rangle$ and the mean square gyration radius $\langle R_g^2 \rangle$, at various $W\!i$, predicted by the multi-chain BD algorithm, with the results of single-chain BD simulations in the dilute limit. Excluded volume interactions are taken into account but hydrodynamic interactions are switched off. The parameter values are as indicated in the figure legend.}
\end{figure*}

Figure~\ref{transshear} displays the growth in the transient viscosity upon inception of steady planar shear flow as a function of time. Clearly, there is excellent agreement between the multi-chain and single chain simulations (for which the neighbour-list and PBCs are not required). In particular, the well known overshoot phenomena in such flows~\cite{birdetal1}, that occurs at high shear rates, is accurately captured by the multi-chain simulations. 

Figure~\ref{Re2Rg2PSFDilVal} displays the dependence of $\langle R_e^2 \rangle$ and $\langle R_g^2 \rangle$ on Weissenberg number $W\!i$ at steady state. Unlike in the Rouse model, since excluded volume interactions are present, analytical expressions for the mean size of the chain, and the zero shear rate viscosity are not known. However, the following relationship between the zero shear rate viscosity and the radius of gyration can be derived in the absence of hydrodynamic interactions, by developing a retarded motion expansion for the stress tensor~\cite{prakash:macromol-01},
\begin{equation} 
\label{etarg} 
\eta_{0} =  \frac{2}{3} \, N_{b} \, \langle R_g^2 \rangle
\end{equation}
This enables the calculation of the relaxation time $\lambda_{\eta}$ once $\langle R_g^2 \rangle$ is known, without the need to estimate $\eta_{0}$ by extrapolating finite shear rate results for $\eta$ to the zero shear rate limit. In this instance, $\lambda_{\eta}=1.184$. Clearly, there is excellent agreement between the multi-chain and single-chain BD simulation results. 

\begin{figure}[!tbp]
\centering
\resizebox{8cm}{!} {\includegraphics{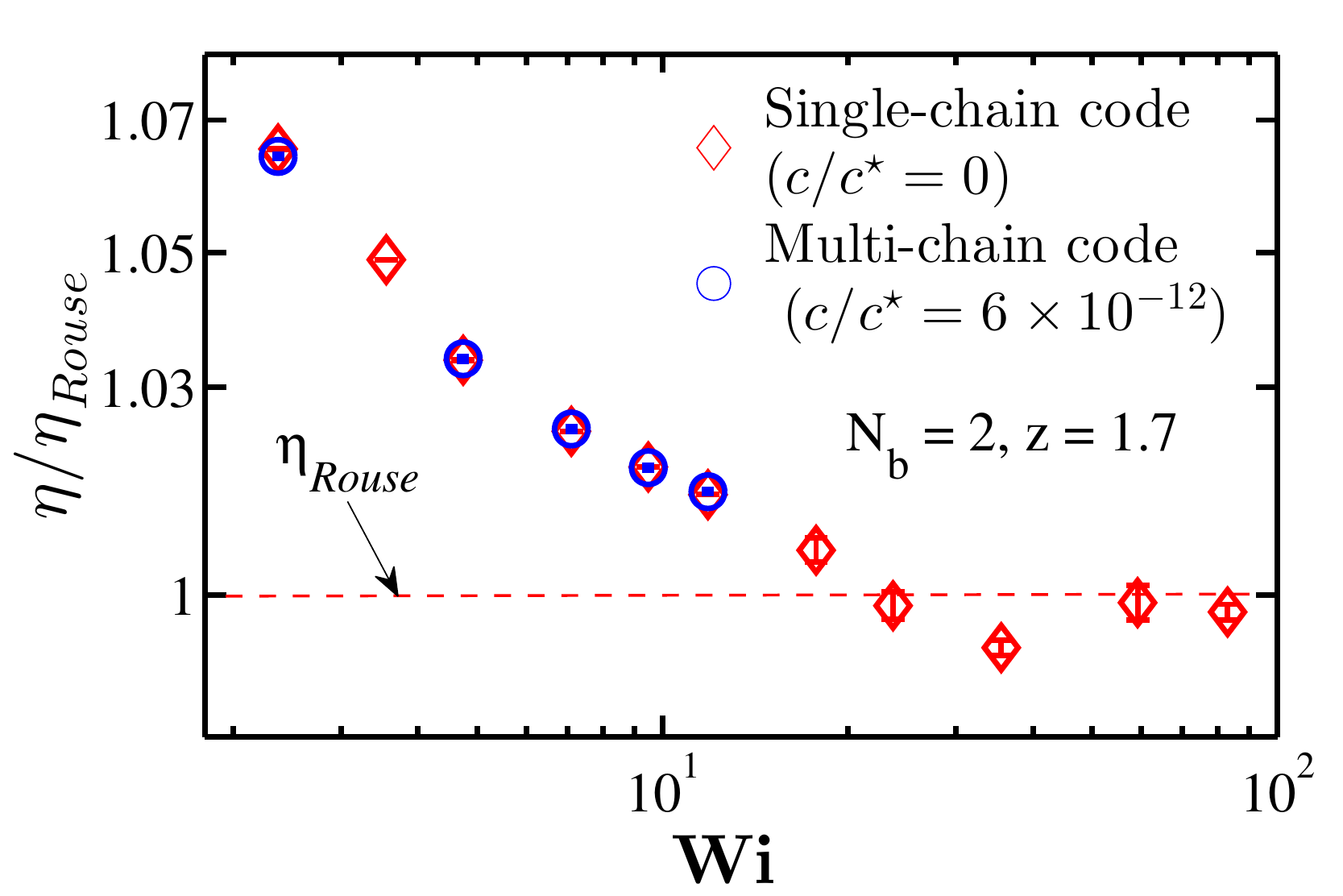}}
\caption{Comparison of the viscosity $\eta$, at various $W\!i$, predicted by the multi-chain BD algorithm with the results of single-chain BD simulations in the ultradilute limit.}
\label{FlowCurveDumbbell}
\end{figure}

Figure~\ref{FlowCurveDumbbell} compares the steady state viscosity ratio $\eta/\eta_\text{Rouse}$ (computed using Eqs.~(\ref{eq:etap-shear}), and~(\ref{eq:rouseeta})), as a function of $W\!i$, predicted by the multi-chain and single chain BD simulations. As is well known, the incorporation of excluded volume interactions into kinetic theory models of polymer solutions leads to the prediction of shear thinning~\cite{Prabhakar2002,kumar:jcp-04}. This is believed to arise for the following reason.  The value of the zero shear rate viscosity is greater in the presence of excluded volume interactions than in its absence,  because of the swelling of the polymer coil. When flow is switched on, however, the increase in the separation between segments of the chain leads to a weakening of excluded volume interactions, and consequently a decrease in the viscosity. The behaviour displayed in Fig.~\ref{FlowCurveDumbbell}  is in line with this expectation, with the viscosity decreasing from its enhanced value at low shear rates, where excluded volume interactions are still strong, to the Rouse viscosity in the limit of high shear rates, where excluded volume interactions are absent. 
Once again, the agreement between multi-chain and single chain simulations indicates the robustness of the former algorithm.

\begin{figure}[!tbp]
\centering
\resizebox{7cm}{!} {\includegraphics{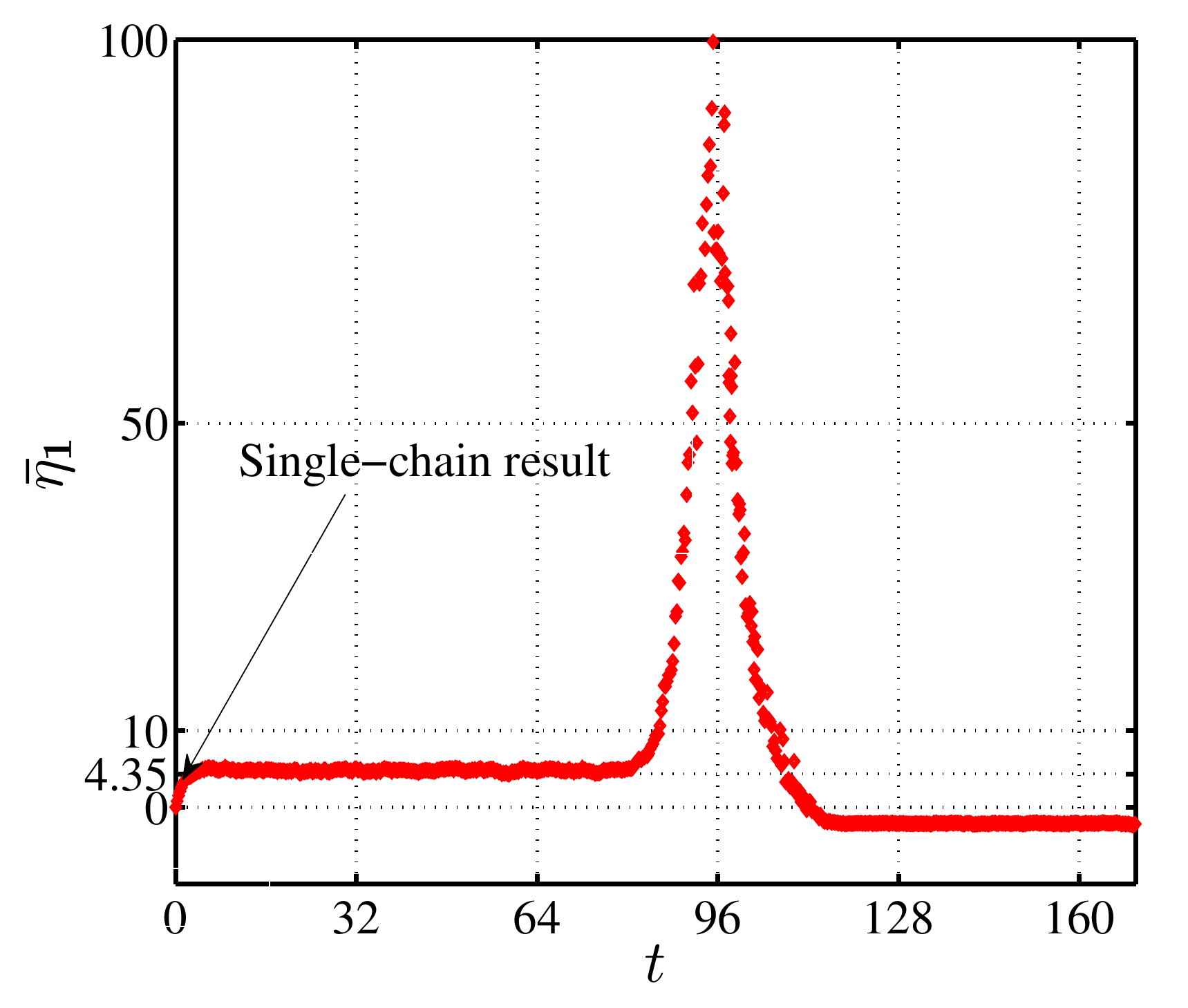}}
\caption{Illustration of numerical instability in planar elongational flow simulations as a result of a naive implementation of the algorithm to evaluate intra-chain bonded interactions.}
\label{Instability}
\end{figure}

We turn our attention now to validation studies for planar elongational flows. PEF simulations provide an opportunity to discuss an aspect of the current implementation of the multi-chain algorithm that differs from the usual practise in NEMD. Essentially, at every time step, there is a need to evaluate forces due to bonded interactions between the beads of a chain, and non-bonded interactions between all the beads in the system. In the course of a simulation, after sufficiently long time, inevitably some of the chains leave the original simulation box because of the action of flow. In NEMD simulations, at every time step, the centre of mass of such chains is mapped back into the original simulation box using the appropriate PBCs, and forces due to both bonded and non-bonded interactions are then calculated. However, in our BD simulations we find that this procedure is not always sufficient, since if the chains are long (which is more common in BD simulations), some parts of the chains may still remain outside the original simulation box after the mapping. For the calculation of forces due to non-bonded interactions, we resolve this problem by mapping all beads that belong to segments of chains that lie outside the original simulation box, using the appropriate PBCs, back into the simulation box. This procedure, however, does not work for the calculation of forces due to bonded interactions since in this case we need to consider chains in their entirety. Using the periodic image within the box, of segments that lie outside the simulation box, leads to an inaccurate calculation of forces due to bonded interactions. Note that this problem is not relevant for non-bonded interactions, since they are always calculated pair-wise between every bead in the simulation box, and every other bead, both in the box and in all the periodic images. A naive alternative is to simply keep track of the absolute positions of the $N$ beads that were in the original simulation box, as a function of time, and to evaluate forces due to intra-chain bonded interactions at every time step. However, in PEF simulations, since the numerical value of the $x$-coordinate of beads increases continuously due to elongation in the $x$-direction, such an implementation leads to numerical instability after a sufficiently long time. This is illustrated in Fig.~\ref{Instability}, which displays the extensional viscosity ${\bar \eta}_1$ for a solution of  FENE dumbbells with finite extensibility parameter $b = 50$, at an elongation rate $\dot{\epsilon} = 0.3$, as a function of time. The correct value of ${\bar \eta}_1$ for these parameters can be shown, by carrying out single-chain BD simulations, to be $4.351 \pm 0.002$. It is clear from Fig.~\ref{Instability}, that the value of ${\bar \eta}_1$ in the multi-chain simulations reaches $4.35$ very rapidly. However, after about $25$ strain periods, a catastrophic change is observed with ${\bar \eta}_1$, settling eventually to a wrong value. 

%
\begin{figure}[tbp]
\begin{center}
\resizebox{7.5cm}{!} {\includegraphics*{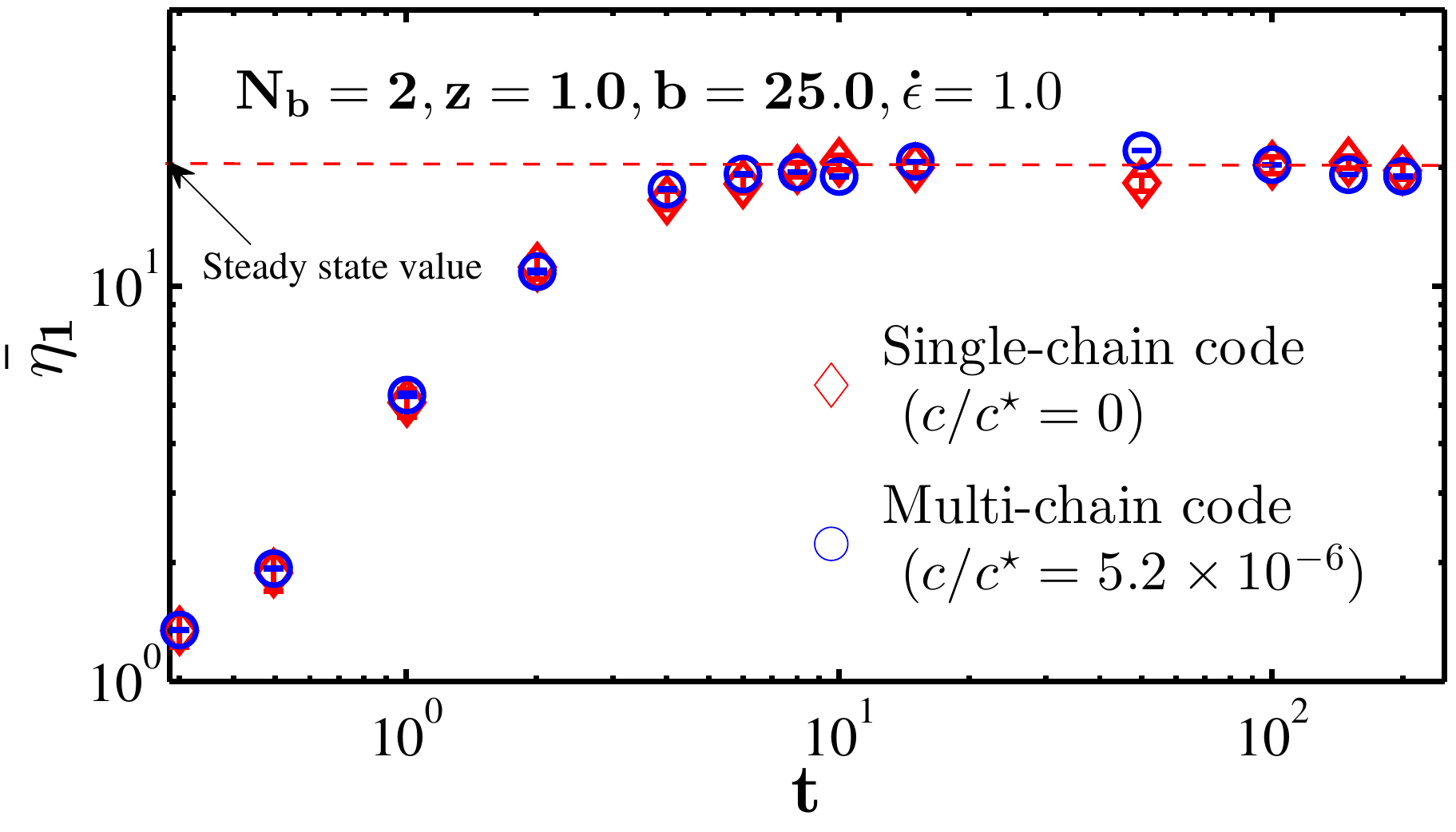}} 
\end{center}
\vskip-20pt
\caption{\small \label{transelong} Comparison of the transient viscosity upon inception of steady planar extensional flow, at non-dimensional times $t$, predicted by the multi-chain BD algorithm, with the results of single-chain BD simulations in the dilute limit. }
\end{figure}

We have adopted the following alternative procedure for calculating the forces due to bonded interactions. After each strain period, we check to see if a chain is in close proximity of the simulation box, based on whether $\vert r_{\nu, x} \vert < f_1 L_1$, $\vert r_{\nu, y} \vert < f_2 L_2$, or $\vert r_{\nu, z} \vert < f_3 L_3$, where, $r_{\nu,x}$, $r_{\nu,y}$ and $r_{\nu,z}$ are the coordinates of bead $\nu$, and $L_1$, $L_2$ and $L_3$ are the magnitudes of cell basis vectors $\Vector{L_1}$, $\Vector{L_2}$ and $\Vector{L_3}$, respectively. The factors $f_1, f_2$ and $f_3$ are arbitrary parameters that are used to set the upper limit on the numerical values of $r_{\nu,x}$, $r_{\nu,y}$ and $r_{\nu,z}$. Here, we set $f_1 = f_2 = f_3 = 2$.  If all the beads of a chain are not in the proximity of the simulation box, then we abandon this chain, and begin to follow the trajectory of the image of this chain that is in the proximity of the simulation box. This does not affect the dynamics of the particles in any way, since we are still tracking the trajectories of the same set of $N$ unique particles and their images. This procedure ensures that the numerical values of the coordinates of the beads never blow up and numerical instability is avoided. Note that the numerical instability observed in NEMD simulations of PEF is not related to the instability discussed here, but is rather due to the lack of momentum conservation that arises from numerical round-off errors \citep{ToddJCP2000}. 

\begin{figure*}[t]
\centering
\begin{tabular}{cc}
\resizebox{7.5cm}{!} {\includegraphics*{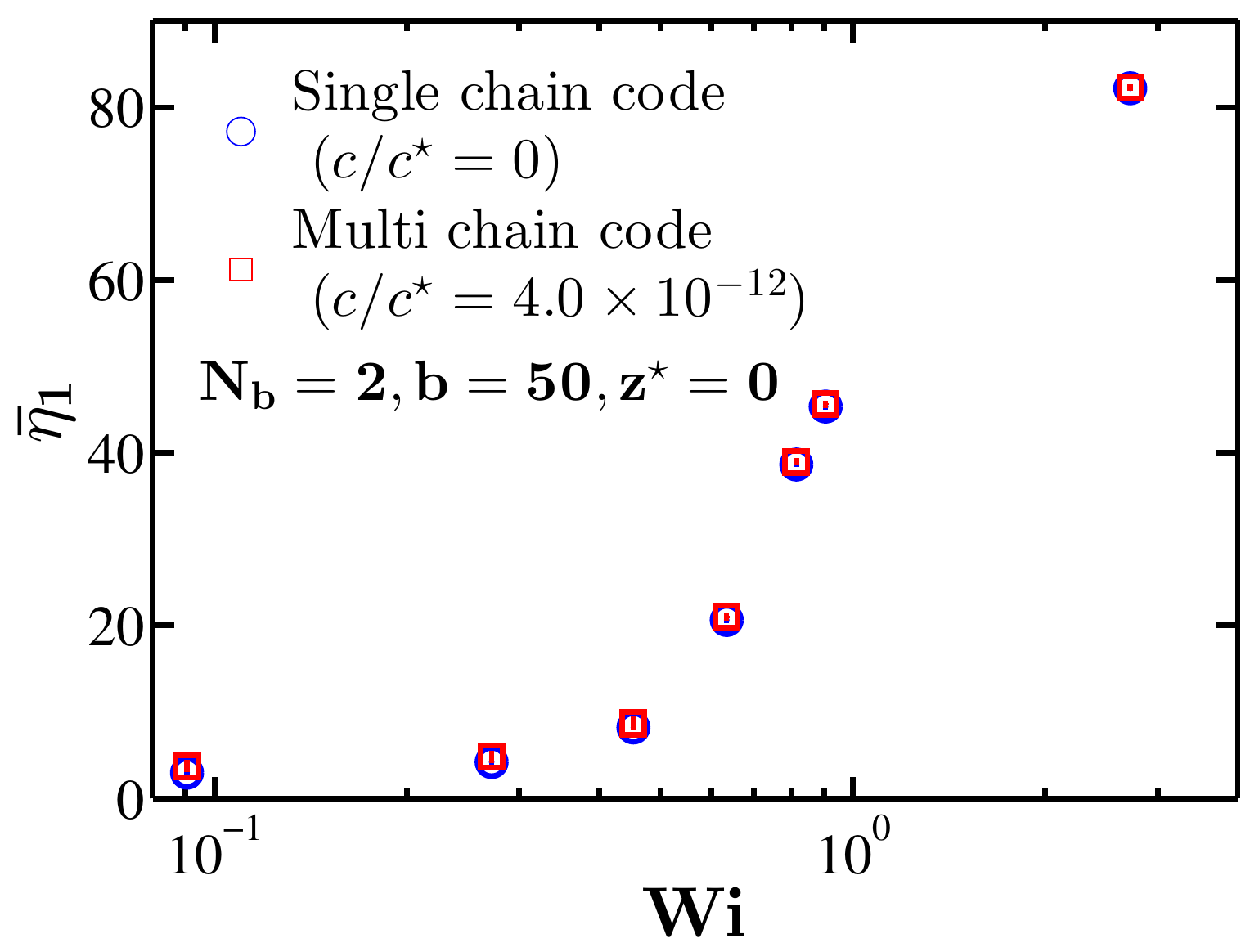}} &
\resizebox{8.5cm}{!} {\includegraphics*{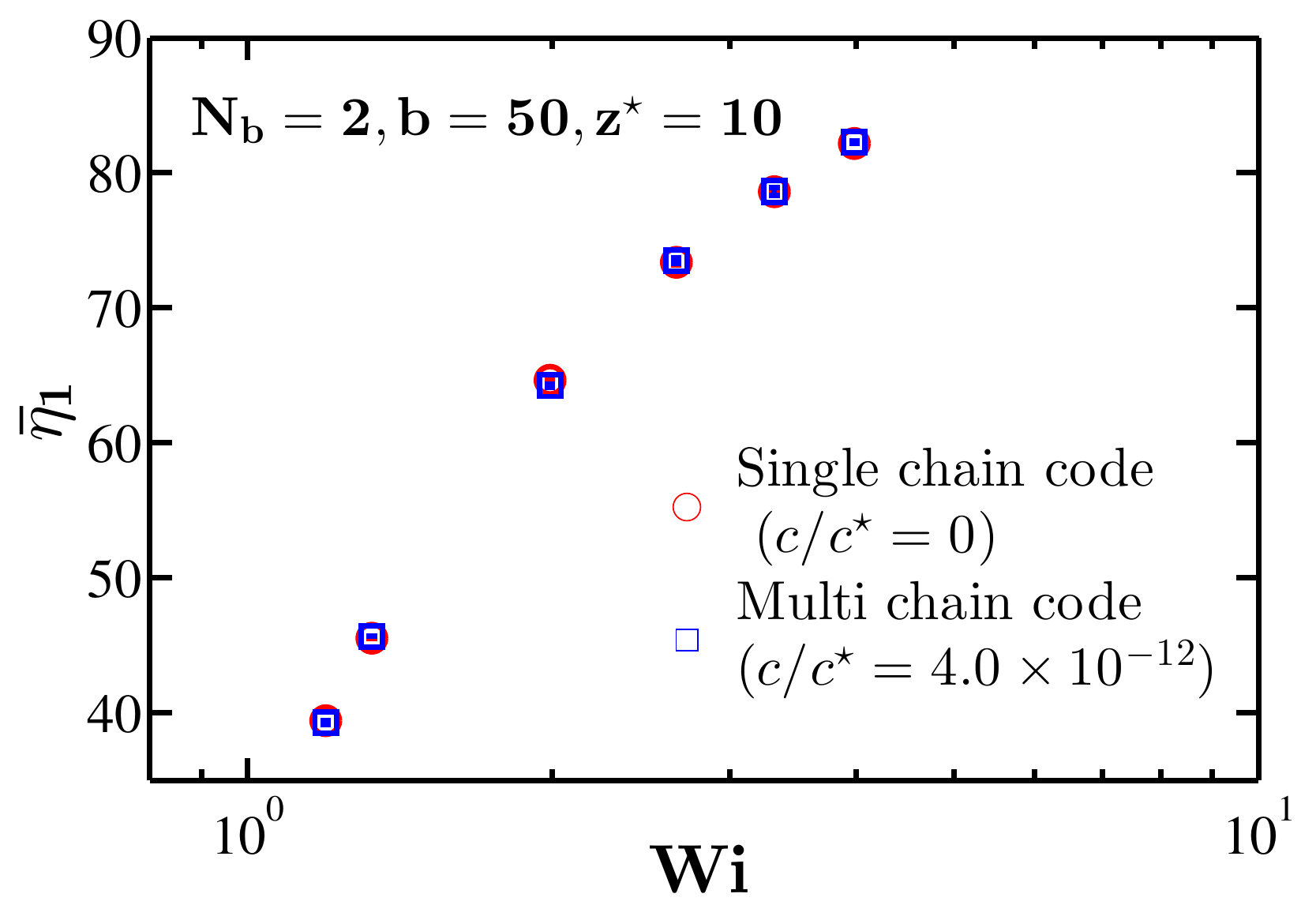}} \\
(a) & (b) \\  
\end{tabular}
\caption{\small \label{PEFValidationEta} Comparison of ${\bar \eta}_1$ predicted by the multi-chain BD algorithm with the results of single-chain BD simulations, at various $\dot{\epsilon}$, in the dilute limit: (a) for $z^\star = 0$ and (b) for $z^\star = 10$.}
\end{figure*}
In all the PEF simulations reported here, we have used $\tilde k = 3$ and $N_{11} = 2$, which are required for calculating the strain period and the magic angle as discussed in the previous section.
For PEF simulations, spring forces cannot be modeled using the Hookean force law, which permits the physically unrealistic indefinite extension of the spring. Since the finite extensibility of the polymer is important in situations where the molecule is likely to be close to full extension, such as in strong shear or elongational flows, a FENE spring force is used here to model spring forces in PEF. 

As in the case of planar shear flow, we first examine the validity of the algorithm for a transient flow, namely, the inception of steady planar elongational flow. Figure~\ref{transelong} displays the growth in the extensional viscosity as a function of time. As is well known, the viscosity increases monotonically as polymer chains unravel from a coiled state to a stretched state under the action of flow, before levelling off to a steady state value~\cite{BirdVol2}. It is clear that the multi-chain algorithm accurately captures the variation of the viscosity with time. The perfect agreement between the multi-chain and single chain simulations in both the transient flows examined here indicates that the remapping of the system at each strain period has been implemented successfully. 

Multi-chain BD simulations have been carried out to obtain the steady state value of ${\bar \eta}_1$ for a range of $\dot{\epsilon}$, for $z^\star = 0$ and for $z^\star = 10$, corresponding to theta and good solvents, respectively. We set $d^\star = 1$, $N_c = 500$, $c/c^\star = 2 \times 10^{-16}$ and the FENE parameter $b = 50$. Simulation results for the two values of $z^\star$ are shown in Figs. \ref{PEFValidationEta} (a) and (b), respectively, obtained by multi-chain and single-chain simulations, in terms the Weissenberg number, which in this case is defined by the expression, $W\!i = \lambda_{\eta} \dot \epsilon$. Clearly, in both cases, there is excellent agreement between the multi-chain and single-chain results, validating the implementation of the current BD algorithm in planar extensional flows. 

\section{Planar mixed flows of polymer solutions at finite concentrations}
\label{Sec:Results}

In this section, we describe the new results of this work, namely, the prediction of polymer size and viscosity in planar mixed flows at finite concentrations. We consider a simple system of FENE dumbbells $(N_{b} = 2)$ with finite extensibility parameter $b = 25$. The excluded volume parameters are chosen to be $z^{\star} = 1/\sqrt{2}$, and $d^{\star} = 0.93$. Data is presented for two values of $c/c^{\star}$: (i) $c/c^{\star} = 0.176$, and (ii) $c/c^{\star} = 1.0$.  The lattice parameters $\tilde k$ and $N_{11}$ are chosen to be $3$ and $2$, respectively, for all the results reported in this section. 

\begin{figure*}[!htbp]
\centering
\begin{tabular}{cc}
\resizebox{8.2cm}{!} {\includegraphics*{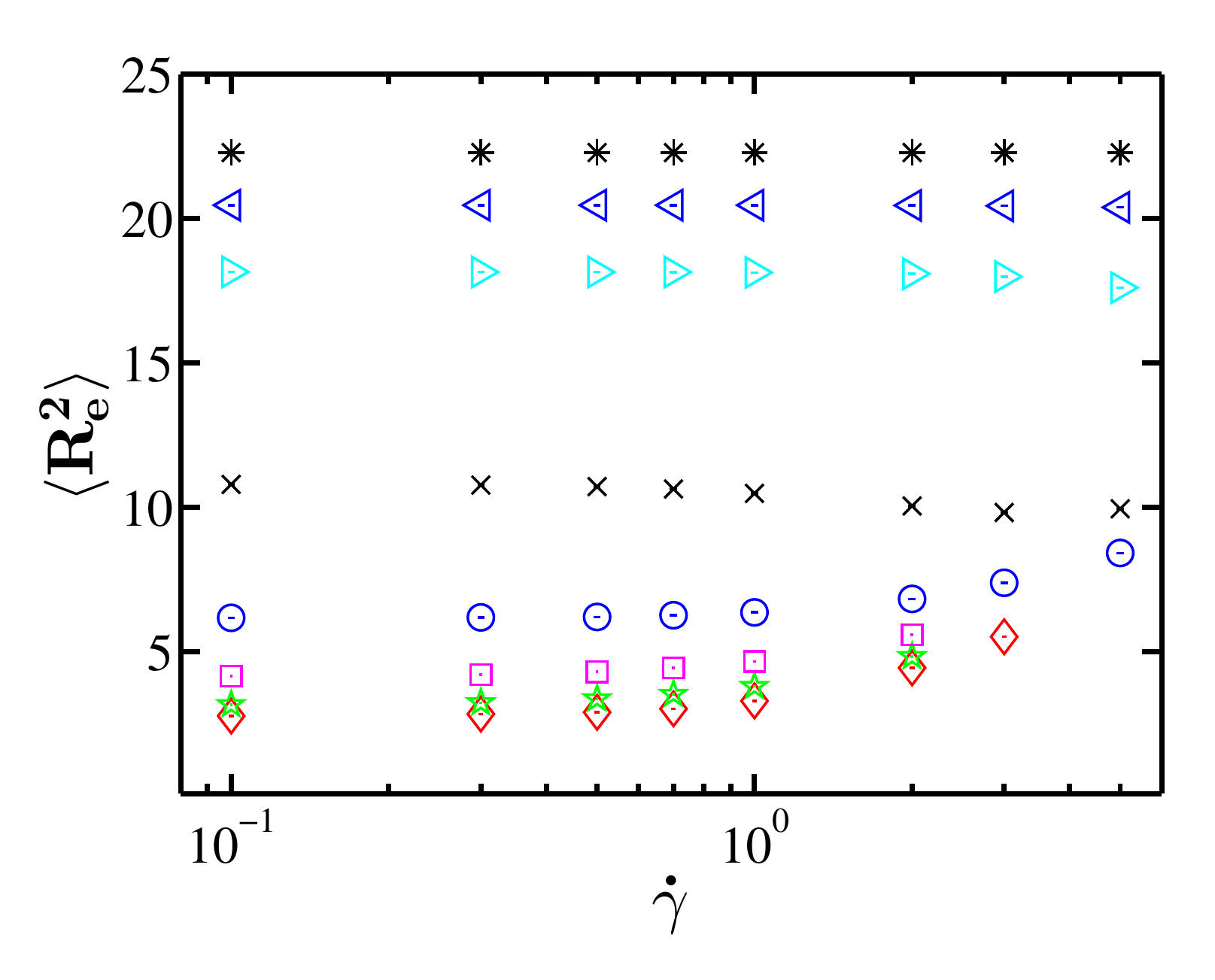}}&
\resizebox{7.7cm}{!} {\includegraphics*{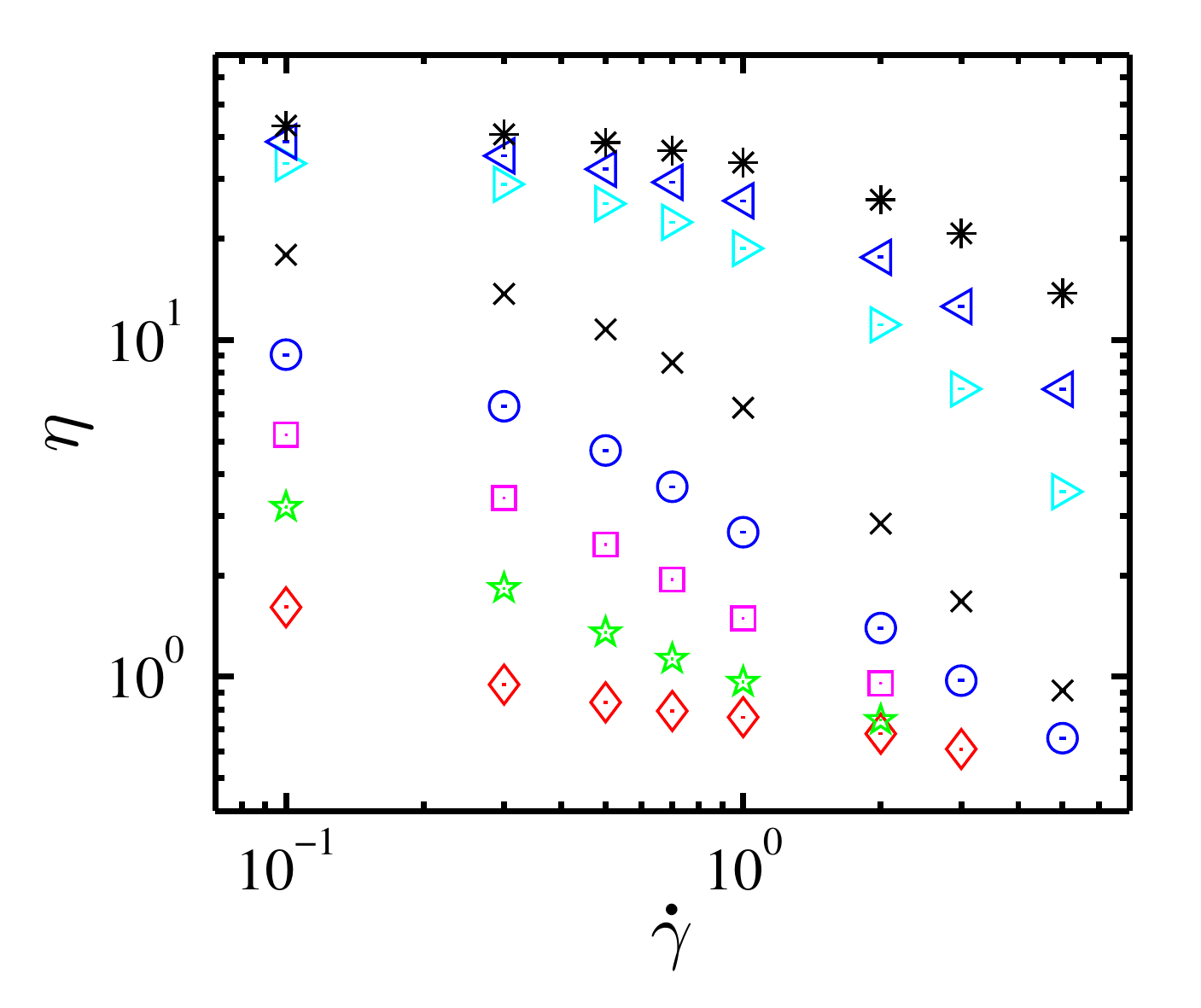}}\\
(a) & (b)  \\
\resizebox{8.2cm}{!} {\includegraphics*{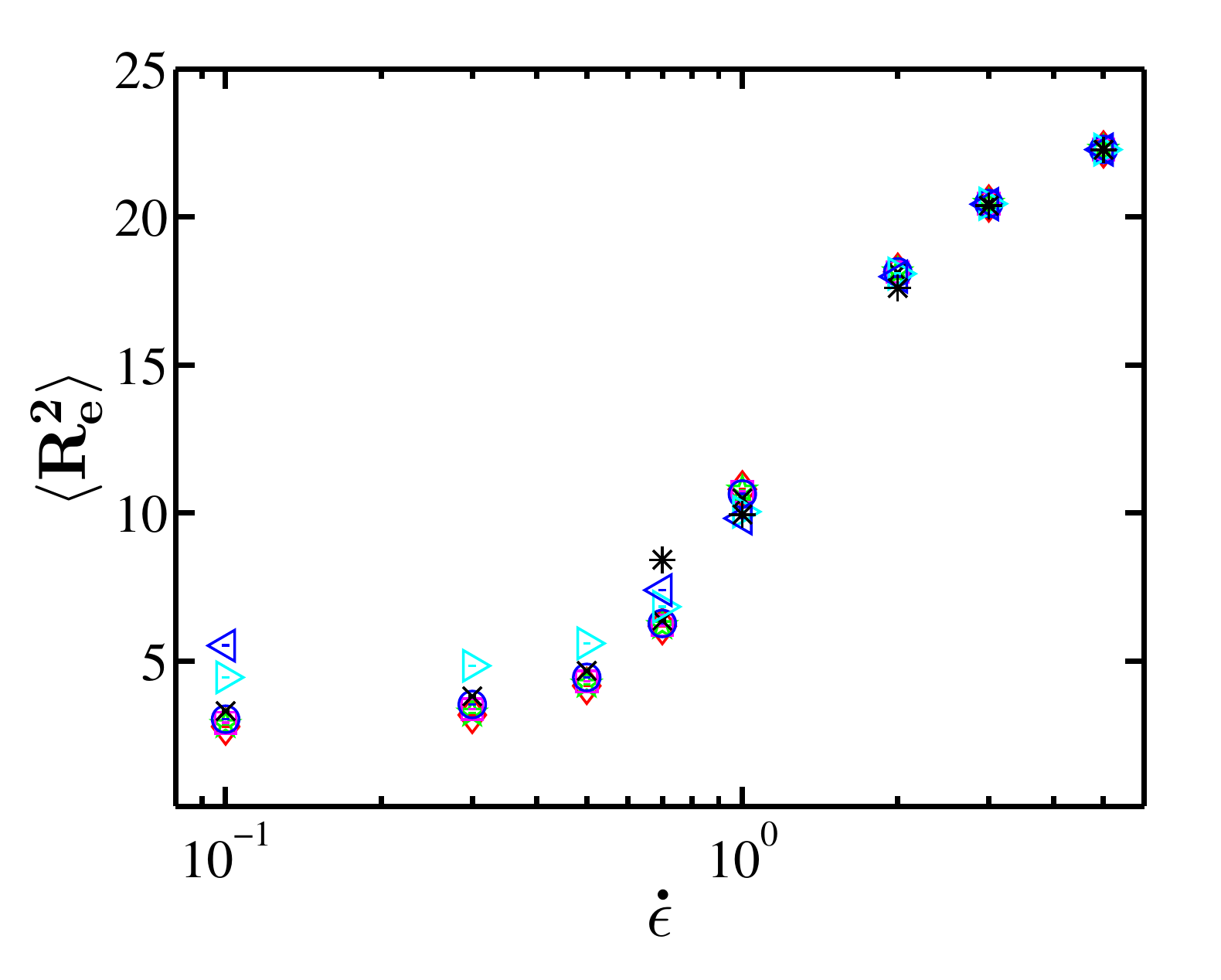}}&
\resizebox{7.7cm}{!} {\includegraphics*{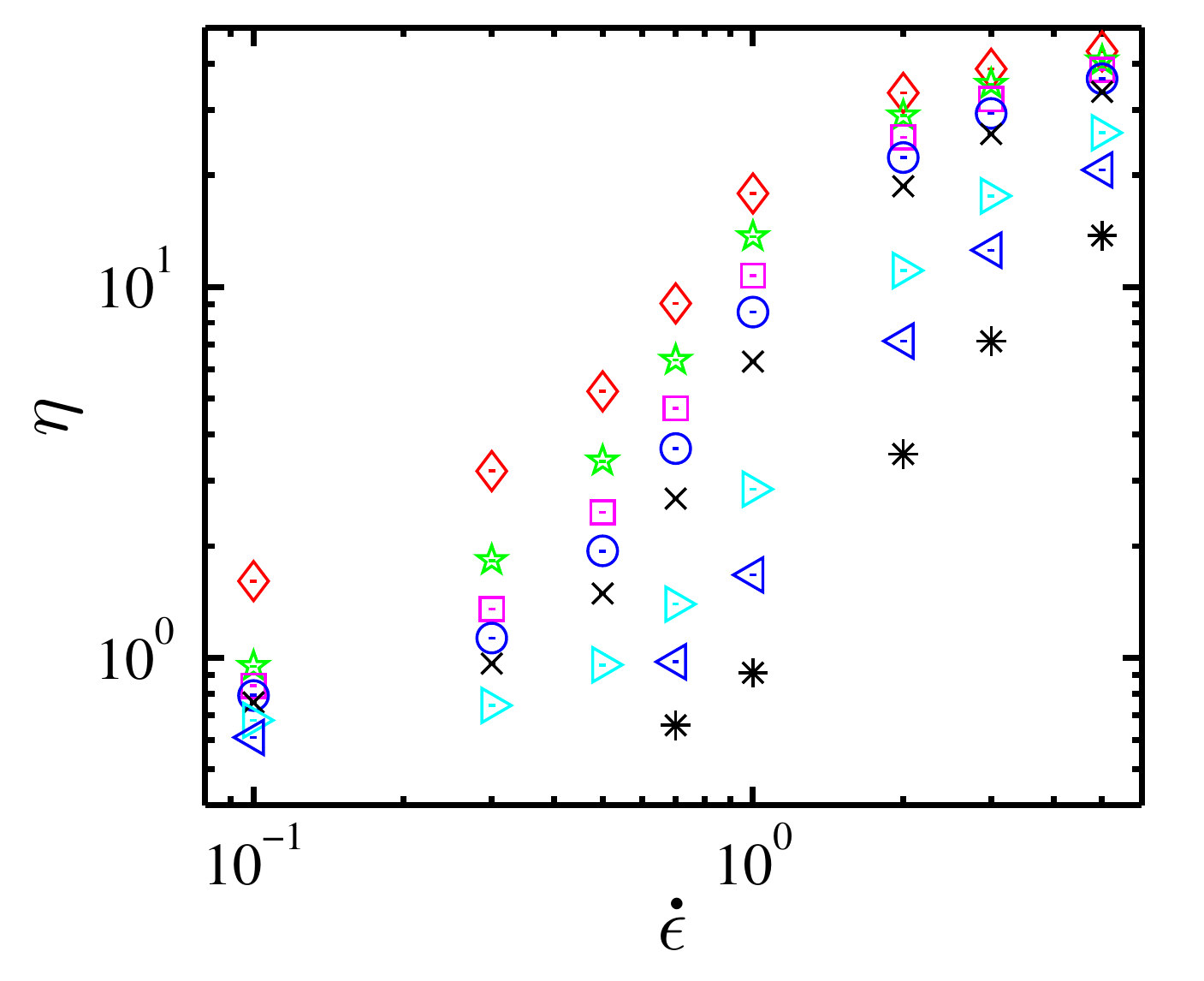}}\\
(c) & (d)  
\end{tabular}
\caption{\label{SizeVisVar1}  Variation of polymer size and viscosity with shear rate $\dot{\gamma}$, and elongation rate $\dot{\epsilon}$, in planar mixed flow.
 (a) Variation of polymer size with $\dot{\gamma}$ at various fixed values of $\dot{\epsilon}$ :  ${\color{red}{\diamond}}$  $\dot{\epsilon} = 0.1$, $\color{green}{\star}$ $\dot{\epsilon} = 0.3$, $\color{magenta}{\Box}$ $\dot{\epsilon} = 0.5$, $\color{blue}{\circ}$ $\dot{\epsilon} = 0.7$, $\color{black}{\times}$ $\dot{\epsilon} = 1.0$, $\color{cyan}{\triangleright}$ $\dot{\epsilon} = 2.0$, $\color{blue}{\triangleleft}$ $\dot{\epsilon} = 3.0$,  $\color{black}{\ast}$ $\dot{\epsilon} = 5.0$; 
(b) Variation of viscosity with $\dot{\gamma}$ at various fixed values of $\dot{\epsilon}$: ${\color{red}{\diamond}}$  $\dot{\epsilon} = 0.1$, $\color{green}{\star}$ $\dot{\epsilon} = 0.3$, $\color{magenta}{\Box}$ $\dot{\epsilon} = 0.5$, $\color{blue}{\circ}$ $\dot{\epsilon} = 0.7$, $\color{black}{\times}$ $\dot{\epsilon} = 1.0$, $\color{cyan}{\triangleright}$ $\dot{\epsilon} = 2.0$, $\color{blue}{\triangleleft}$ $\dot{\epsilon} = 3.0$,  $\color{black}{\ast}$ $\dot{\epsilon} = 5.0$; 
 (c) Variation of polymer size with $\dot{\epsilon}$ at various fixed values of $\dot{\gamma}$ : ${\color{red}{\diamond}}$  $\dot{\gamma} = 0.1$, $\color{green}{\star}$ $\dot{\gamma} = 0.3$, $\color{magenta}{\Box}$ $\dot{\gamma} = 0.5$, $\color{blue}{\circ}$ $\dot{\gamma} = 0.7$, $\color{black}{\times}$ $\dot{\gamma} = 1.0$, $\color{cyan}{\triangleright}$ $\dot{\gamma} = 2.0$, $\color{blue}{\triangleleft}$ $\dot{\gamma} = 3.0$,  $\color{black}{\ast}$ $\dot{\gamma} = 5.0.$;
 (d) Variation of viscosity with $\dot{\epsilon}$ at various fixed values of $\dot{\gamma}$: ${\color{red}{\diamond}}$  $\dot{\gamma} = 0.1$, $\color{green}{\star}$ $\dot{\gamma} = 0.3$, $\color{magenta}{\Box}$ $\dot{\gamma} = 0.5$, $\color{blue}{\circ}$ $\dot{\gamma} = 0.7$, $\color{black}{\times}$ $\dot{\gamma} = 1.0$, $\color{cyan}{\triangleright}$ $\dot{\gamma} = 2.0$, $\color{blue}{\triangleleft}$ $\dot{\gamma} = 3.0$,  $\color{black}{\ast}$ $\dot{\gamma} = 5.0$. In these simulations, $N_{b} = 2$, $b = 25$, $z = 1$, $d^{\star} = 0.93$, and $c/c^{\star} = 0.176$. }  
 \end{figure*}

The influence of shear rate $\dot{\gamma}$  on the polymer size and viscosity at a fixed value of elongation rate $\dot{\epsilon}$, is examined in Figs.~\ref{SizeVisVar1}~(a) and~(b), while the influence of elongation rate $\dot{\epsilon}$  at a fixed value of shear rate $\dot{\gamma}$,  is examined in Figs.~\ref{SizeVisVar1}~(c) and~(d). There are several features that can be discerned from these figures, which we discuss in turn below. 

\begin{figure*}[!htbp]
\centering
\begin{tabular}{cc}
\resizebox{7.7cm}{!} {\includegraphics*{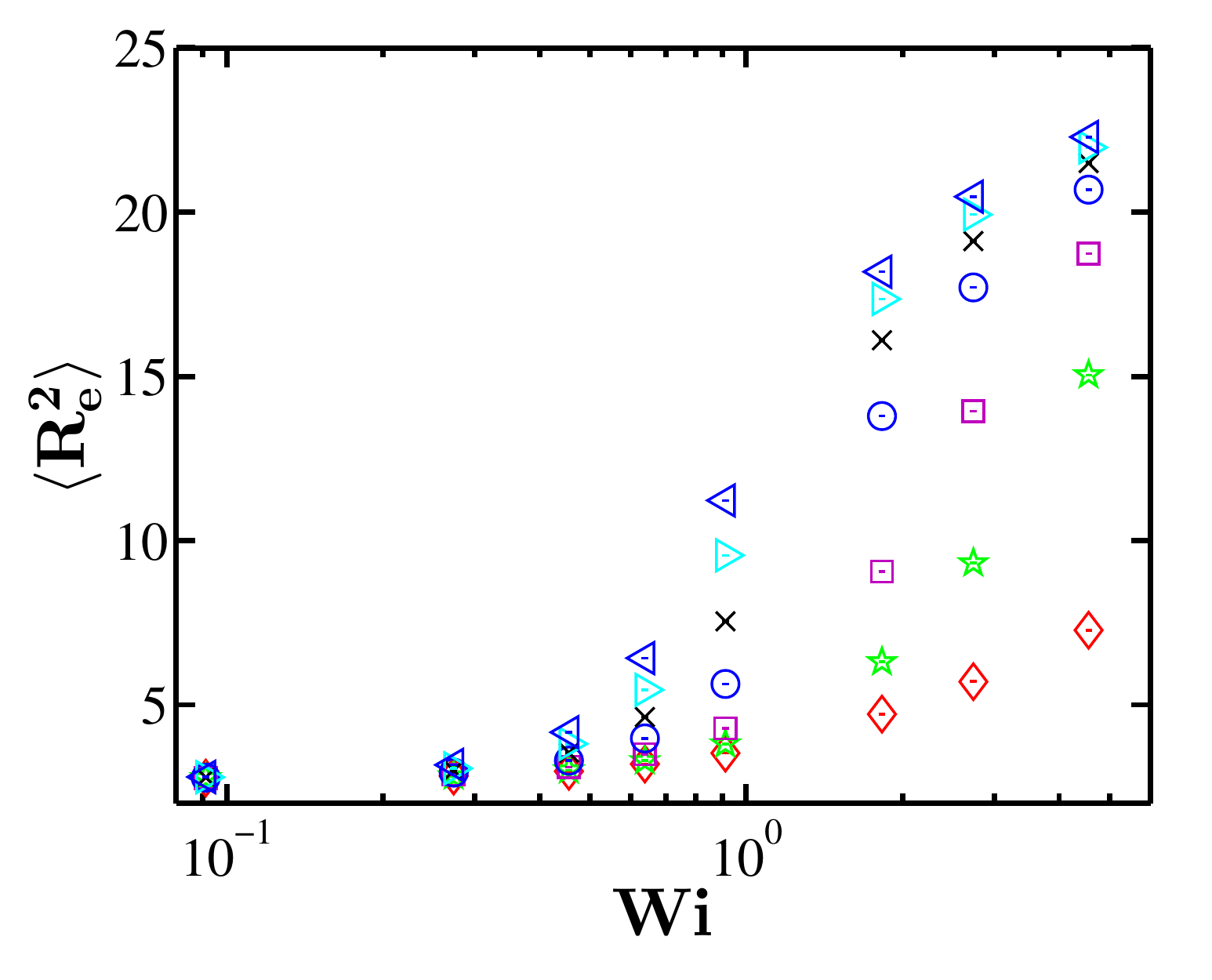}}&
\resizebox{7.8cm}{!} {\includegraphics*{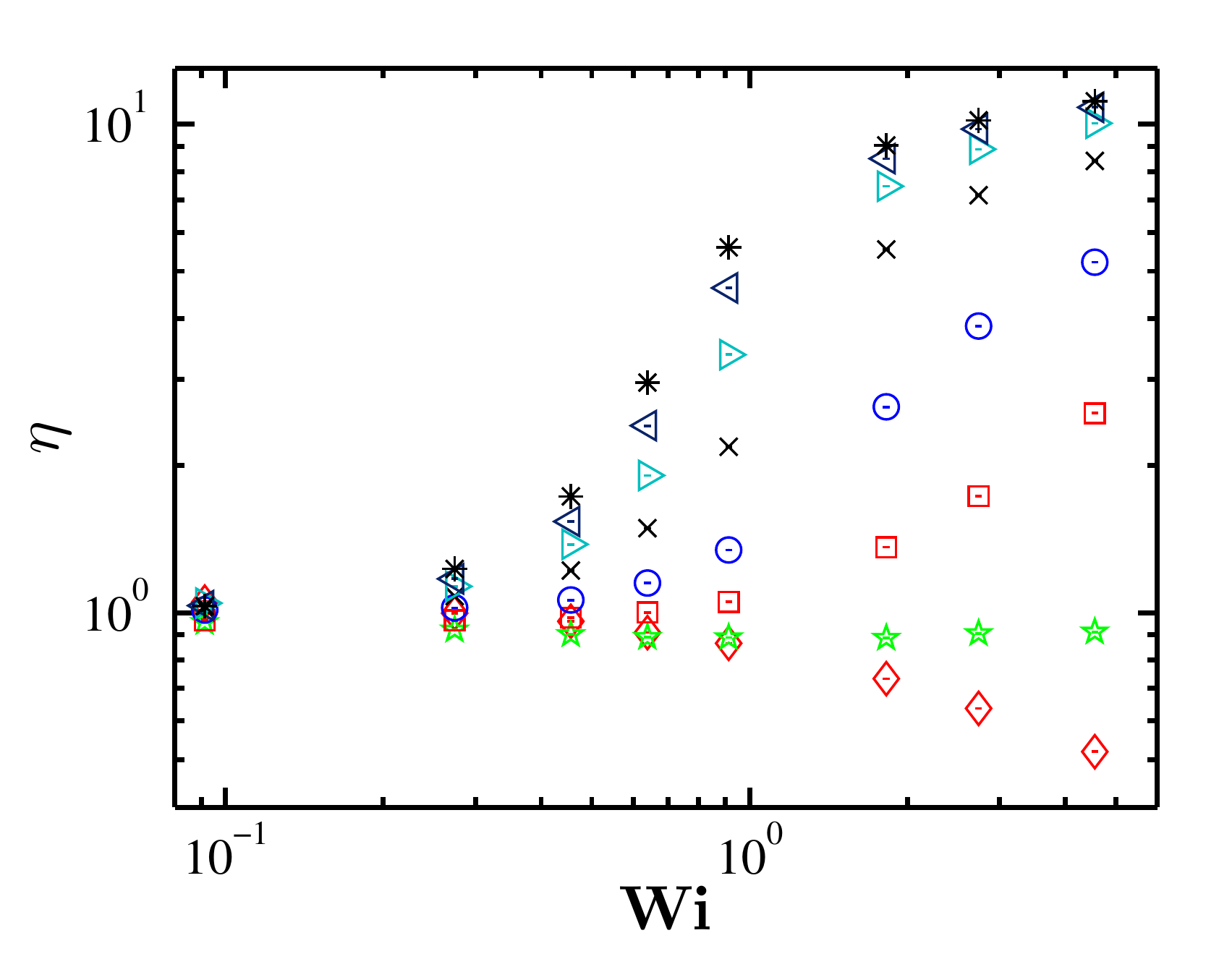}}\\
(a) & (b)  \\
\resizebox{8.2cm}{!} {\includegraphics*{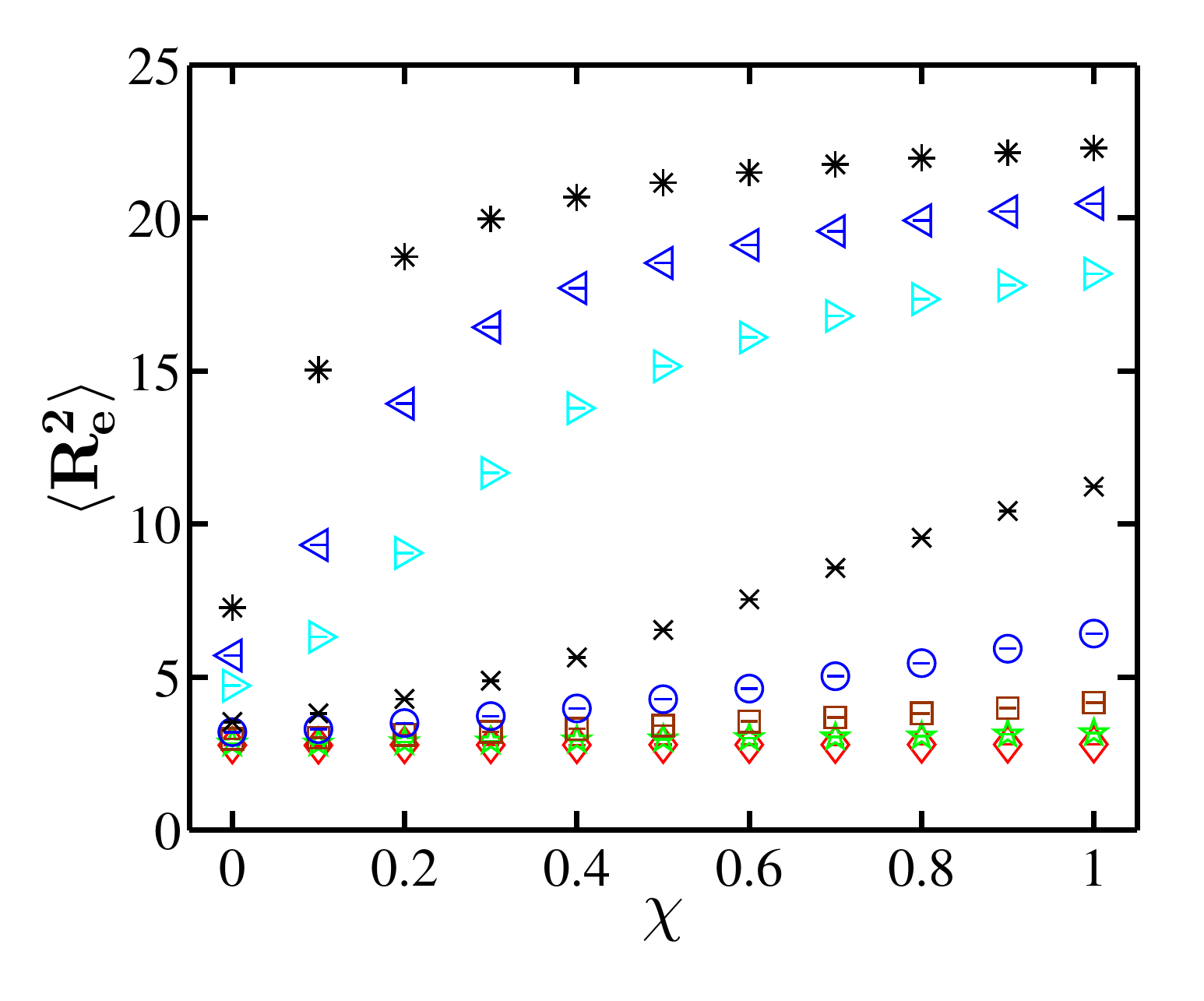}}&
\resizebox{7.6cm}{!} {\includegraphics*{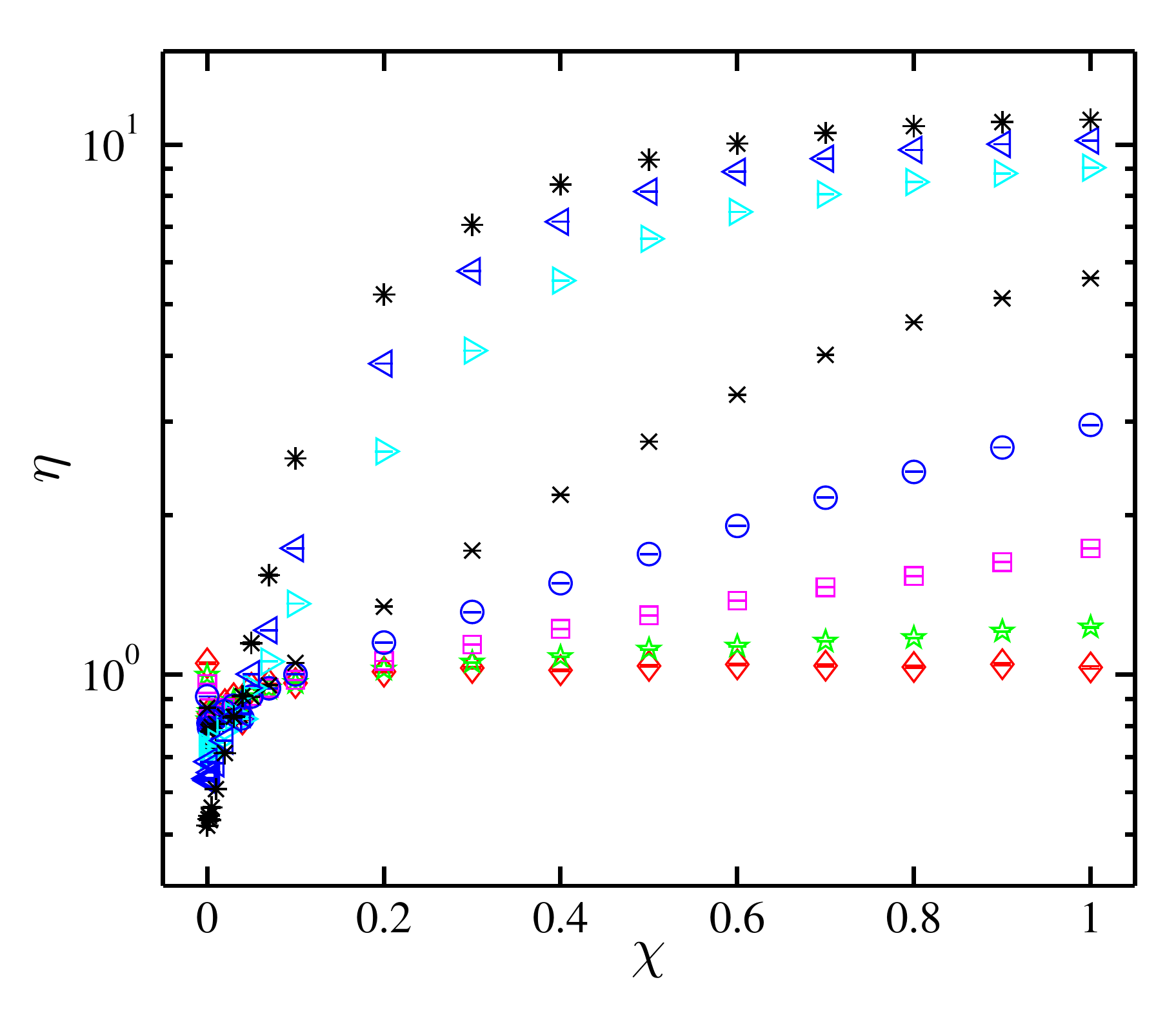}}\\
(c) & (d)  
\end{tabular}
\caption{\label{SizeVisVar2}  Variation of polymer size and viscosity with flow strength $\dot{\Gamma}$, and mixedness parameter $\chi$, in planar mixed flow.  
(a) Variation of polymer size with $\dot{\Gamma}$ at various fixed values of $\chi$: ${\color{red}{\diamond}}$  $\chi = 0.0$, $\color{green}{\star}$ $\chi = 0.1$, $\color{magenta}{\Box}$ $\chi = 0.2$, $\color{blue}{\circ}$ $\chi = 0.4$, $\color{black}{\times}$ $\chi = 0.6$, $\color{cyan}{\triangleright}$ $\chi = 0.8$, $\color{blue}{\triangleleft}$ $\chi = 1.0$;
(b) Variation of viscosity with $\dot{\Gamma}$ at various fixed values of $\chi$: ${\color{red}{\diamond}}$  $\chi = 0.0$, $\color{green}{\star}$ $\chi = 0.04$, $\color{magenta}{\Box}$ $\chi = 0.1$, $\color{blue}{\circ}$ $\chi = 0.2$, $\color{black}{\times}$ $\chi = 0.4$, $\color{cyan}{\triangleright}$ $\chi = 0.6$, $\color{blue}{\triangleleft}$ $\chi = 0.8$,  $\color{black}{\ast}$ $\chi = 1.0$;
(c) Variation of polymer size with $\chi$ at various fixed values of $\dot{\Gamma}$: ${\color{red}{\diamond}}$  $\dot{\Gamma} = 0.1$, $\color{green}{\star}$ $\dot{\Gamma} = 0.3$, $\color{magenta}{\Box}$ $\dot{\Gamma} = 0.5$, $\color{blue}{\circ}$ $\dot{\Gamma} = 0.7$, $\color{black}{\times}$ $\dot{\Gamma} = 1.0$, $\color{cyan}{\triangleright}$ $\dot{\Gamma} = 2.0$, $\color{blue}{\triangleleft}$ $\dot{\Gamma} = 3.0$,  $\color{black}{\ast}$ $\dot{\Gamma} = 5.0$;
(d) Variation of viscosity with $\chi$ for various fixed values of $\dot{\Gamma}$: ${\color{red}{\diamond}}$  $\dot{\Gamma} = 0.1$, $\color{green}{\star}$ $\dot{\Gamma} = 0.3$, $\color{magenta}{\Box}$ $\dot{\Gamma} = 0.5$, $\color{blue}{\circ}$ $\dot{\Gamma} = 0.7$, $\color{black}{\times}$ $\dot{\Gamma} = 1.0$, $\color{cyan}{\triangleright}$ $\dot{\Gamma} = 2.0$, $\color{blue}{\triangleleft}$ $\dot{\Gamma} = 3.0$,  $\color{black}{\ast}$ $\dot{\Gamma} = 5.0$. In these simulations, $N_{b} = 2$, $b = 25$, $z = 1$, $d^{\star} = 0.93$, and $c/c^{\star} = 0.176$. }
\end{figure*}

Fig.~\ref{SizeVisVar1}~(a) indicates that at any value of shear rate $\dot{\gamma}$, the polymer size increases with increasing elongation rate $\dot{\epsilon}$. This is to be expected since it is well known that chains unravel in extensional flows. The interesting point to note is that for $\dot{\epsilon} < 1$, $\langle R_e^2 \rangle$ increases with increasing $\dot{\gamma}$ until it asymptotes to a value of $\langle R_e^2 \rangle \approx 10$, while for $\dot{\epsilon} \ge 1$, $\langle R_e^2 \rangle$ \emph{decreases} with increasing $\dot{\gamma}$, and appears to be reaching the same asymptotic value. Several experimental and theoretical studies of polymer conformations in pure simple shear flow have shown that the polymer size increases with increasing shear rate and typically saturates to roughly $40 \%$ of its fully stretched size. The chain is never fully stretched in shear flow because it experiences repeated stretching and tumbling events (readers can find an extended discussion of  chain conformations in shear flow in Ref.~\citenum{Dalal2012}, and references therein). This is consistent with the results in Fig.~\ref{SizeVisVar1} (a) for $\dot{\epsilon} < 1$, since the square of the fully stretched contour length in the current simulations is given by the parameter $b = 25$ (in non-dimensional units). The decrease in polymer size with increasing shear rates, for $\dot{\epsilon} \ge 1$, can be understood from the fact that at any given elongation rate $\dot{\epsilon}$, the flow becomes increasingly shear dominated at sufficiently high values of $\dot{\gamma}$. This can be seen from the expression,
\begin{equation}
\label{chigamma}
\frac{\dot{\gamma}}{\dot{\epsilon}} = \frac{\left(1-\chi\right)}{\sqrt{\chi}}
\end{equation}
For a fixed value of $\dot{\epsilon}$, as $\dot{\gamma} \rightarrow \infty$, the mixedness parameter $\chi \rightarrow 0$. As a result, at sufficiently high values of $\dot{\gamma}$, we expect the polymer size to asymptote to its value in pure shear flow, regardless of the value of $\dot{\epsilon}$.

Fig.~\ref{SizeVisVar1}~(b) shows that at any value of shear rate $\dot{\gamma}$, the polymer contribution to the solution viscosity, $\eta$, increases with increasing elongation rate $\dot{\epsilon}$. This behaviour is directly correlated with the size of the chain, since a larger chain size implies a larger volume fraction occupied by the chain, and consequently a larger viscosity. The shear thinning that is evident with increasing $\dot{\gamma}$, at all values of $\dot{\epsilon}$, is because of the inclusion of finite extensibility and excluded volume interactions in the model. The influence of these non-linear mesoscopic phenomena on dilute polymer solution behaviour in pure shear flows, has been discussed in detail in Refs.~\citenum{Prabhakar2002} and~\citenum{PrabhakarJNNFM2004}. 

The unravelling of the polymer chain with increasing $\dot{\epsilon}$, at all values of $\dot{\gamma}$, is clearly evident in Fig.~\ref{SizeVisVar1}~(c). As is well known, in pure extensional flows, the conformation of a chain changes from being coil-like at low extension rates, to being fully stretched and rod-like at high extension rates, undergoing a coil-stretch transition at intermediate extension rates~\cite{RaviKARJ2009}. At the lowest values of $\dot{\epsilon}$, the increase in $\langle R_e^2 \rangle$ with increasing $\dot{\gamma}$ is discernible on the scale of the figure. However, at values of $\dot{\epsilon} \gtrsim 10$, changes in $\dot{\gamma}$ have negligible influence on $\langle R_e^2 \rangle$. From Eq.~(\ref{chigamma}), it is clear that at a fixed value of $\dot{\gamma}$, as $\dot{\epsilon} \rightarrow \infty$, the mixedness parameter $\chi \rightarrow 1$. As a result, at sufficiently high values of $\dot{\epsilon}$, we expect the polymer size to asymptote to its fully stretched value in pure elongational flow, \ie, $\left<R_e^2\right>/b \rightarrow 1$ as $\dot{\epsilon} \rightarrow \infty$, regardless of the value of $\dot{\gamma}$.

The behaviour of the polymer contribution to solution viscosity displayed in Fig.~\ref{SizeVisVar1}~(d), can be understood in the light of the results shown in Figs.~\ref{SizeVisVar1}~(b) and~(c). At any value of extension rate $\dot{\epsilon}$, $\eta$ decreases with increasing shear rate $\dot{\gamma}$, because of shear thinning. However, $\eta$ increases with increasing $\dot{\epsilon}$ at all values of $\dot{\gamma}$, because the chain undergoes a coil-stretch transition in this process. The levelling off of $\eta$ to a constant value at high extension rates, is related to the chain reaching its maximum state of stretch, at that particular value of $\dot{\gamma}$.

A completely different and valuable insight is obtained when we consider the behaviour of $\eta$ as a function of $\dot{\Gamma}$ and $\chi$, instead of $\dot{\gamma}$ and $\dot{\epsilon}$. In contrast to $\eta$, however, the variation of $\left<R_e^2\right>$ with $\dot{\Gamma}$ and $\chi$ does not have many features that cannot be anticipated from the results already displayed in Figs.~\ref{SizeVisVar1}~(a) and~(c). These observations are discussed in greater detail  below in the context of Figs.~\ref{SizeVisVar2}~(a) to~(d), where results are presented in terms of a non-dimensional Weissenberg number defined by the expression, $W\!i = \lambda_{\eta} \dot \Gamma$. 

We anticipate that with increasing flow strength $W\!i$, the polymer size $\left<R_e^2\right>$ will increase, regardless of the value of $\chi$. This is indeed the case, as displayed in  Fig.~\ref{SizeVisVar2}~(a). Since a polymer chain tumbles continuously in shear flow while undergoing exponential stretching in extensional flows, the change in $\left<R_e^2\right>$ will become more pronounced as the value of $\chi$ changes from 0 to 1, over a similar range of values of $W\!i$. This behaviour is evident in Fig.~\ref{SizeVisVar2}~(c).

The behaviour of $\eta$ displayed in Figs.~\ref{SizeVisVar2}~(b) and~(d) demonstrates the existence of a critical value of the mixedness parameter, $\chi_\text{c}$, such that for $\chi < \chi_\text{c}$, the flow is shear dominated, while being extension dominated for values of $\chi > \chi_\text{c}$. For instance, as can be seen from Fig.~\ref{SizeVisVar2}~(b), the viscosity \emph{decreases} with increasing flow strength at $\chi = 0$, while \emph{increasing} with $W\!i$ for all values of $\chi > 0.04$. At $\chi=0.04$, the viscosity appears to be nearly independent of flow strength. The precise value of $\chi_\text{c}$ will be discussed in greater detail shortly below, however, the alteration in the variation of $\eta$ with $W\!i$ can be seen more dramatically in Fig.~\ref{SizeVisVar2}~(d), where the viscosity  appears to be shear thinning for values of $\chi$ close to 0, but extension hardening for all large values of $\chi$. 

\begin{figure*}[]
\centering
\begin{tabular}{c}
\resizebox{8.5cm}{!} {\includegraphics*{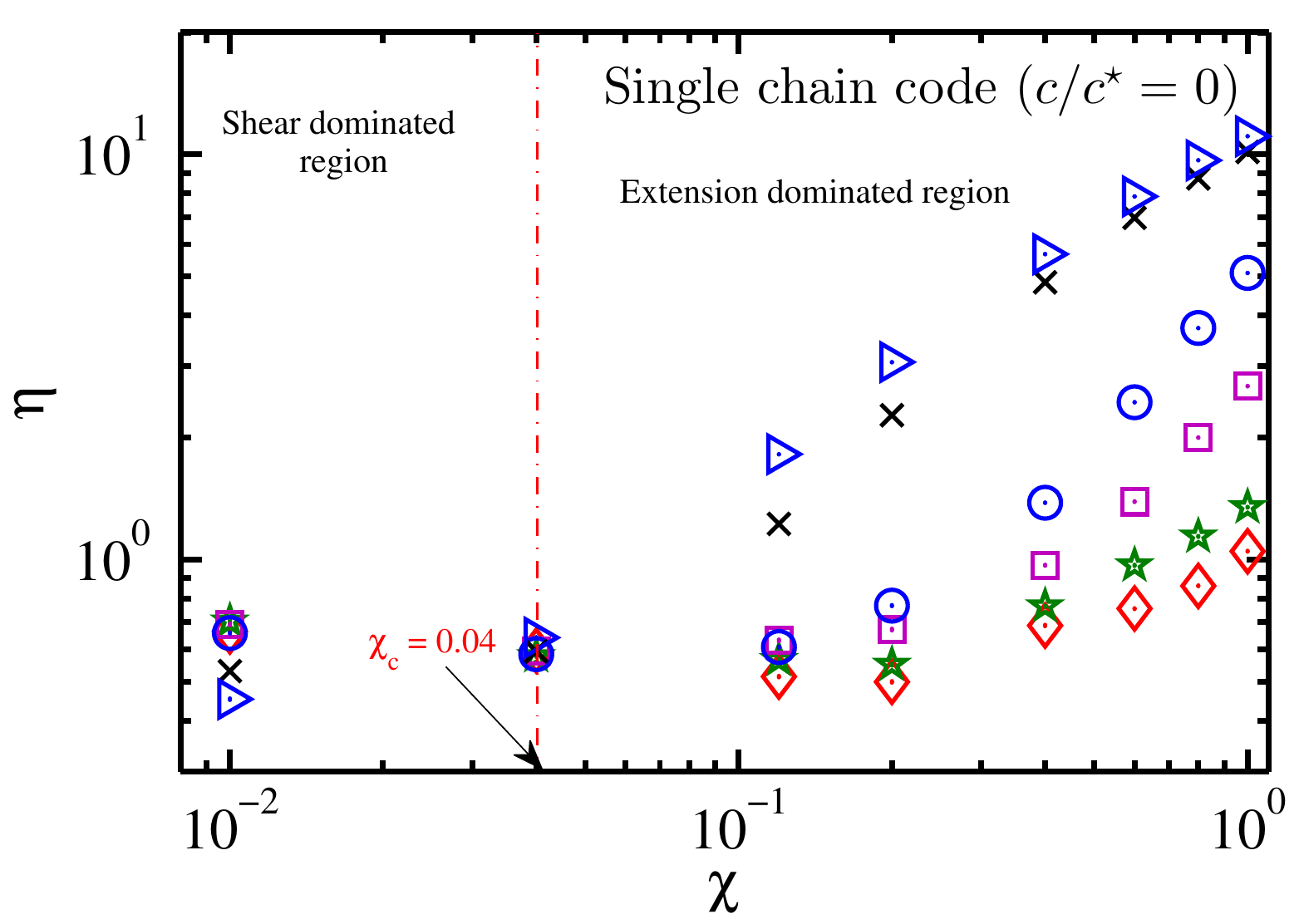}} \\
 (a) \\
\resizebox{8.7cm}{!} {\includegraphics*{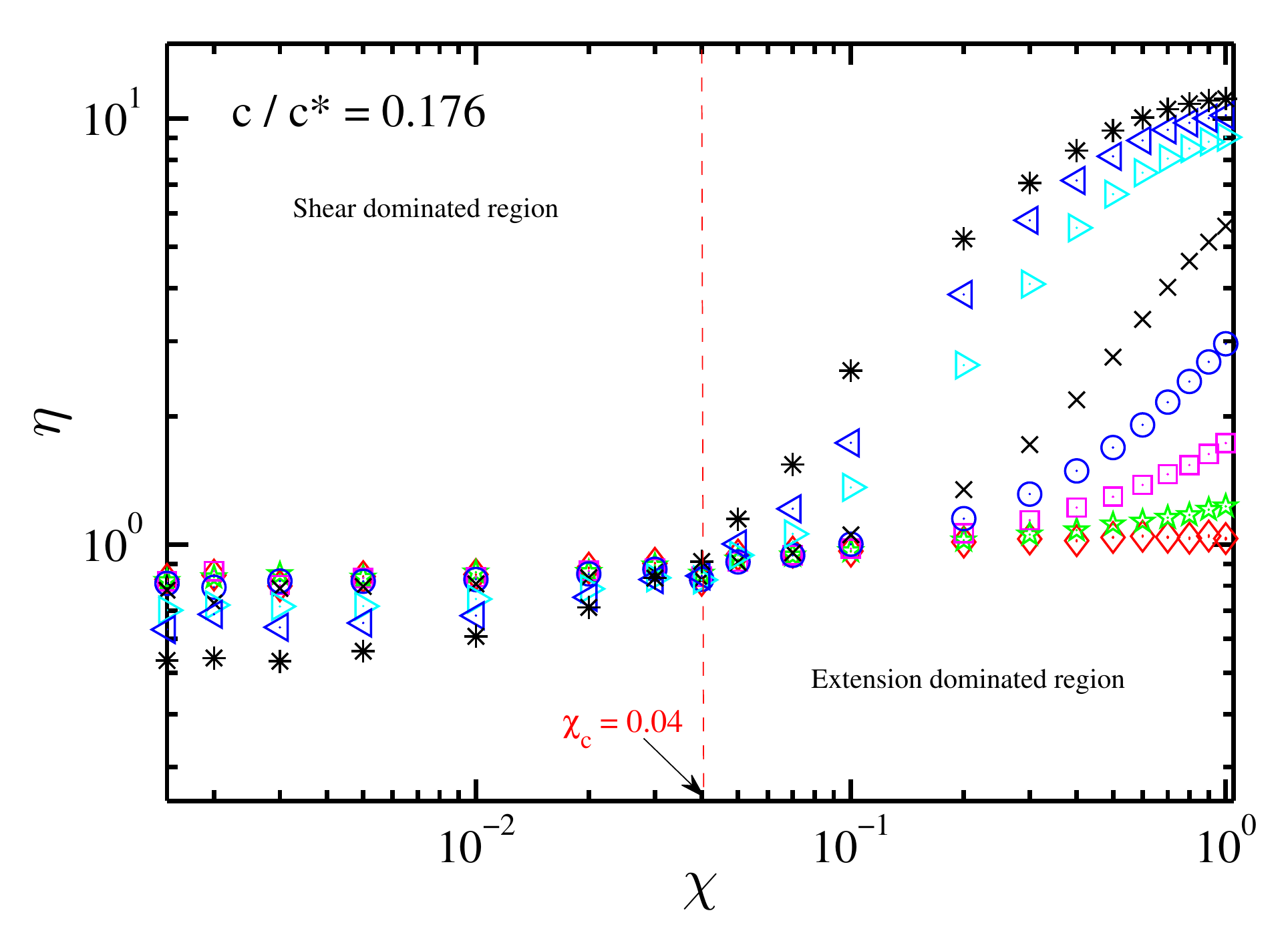}}\\
(b) \\
\resizebox{9.4cm}{!} {\includegraphics*{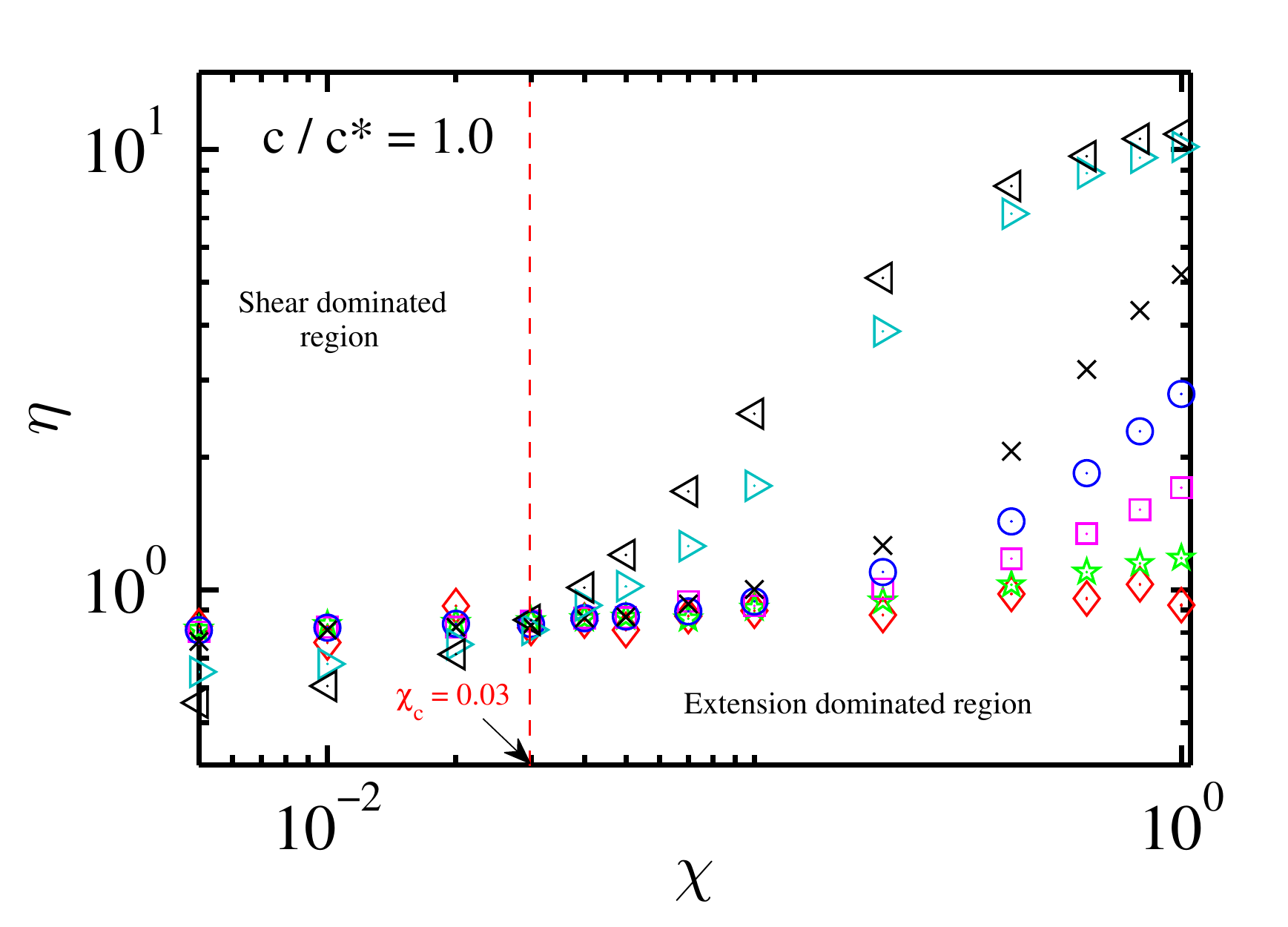}} \\
(c)
\end{tabular}
\caption{\small \label{ChiMixPara} Demonstration of the existence of a critical mixedness parameter, $\chi_\text{c}$, in planar mixed flows. The flow is shear dominated for $\chi < \chi_\text{c}$, while being extension dominated for values of $\chi > \chi_\text{c}$. (a) Variation of $\eta$ with $\chi$ for a dilute solution: ${\color{red}{\diamond}}$  $\dot{\Gamma} = 0.1$, $\color{green}{\star}$ $\dot{\Gamma} = 0.4$, $\color{magenta}{\Box}$ $\dot{\Gamma} = 0.7$, $\color{blue}{\circ}$ $\dot{\Gamma} = 1.0$, $\color{black}{\times}$ $\dot{\Gamma} = 3.0$, $\color{cyan}{\triangleright}$ $\dot{\Gamma} = 5.0$; (b) Variation of $\eta$ with $\chi$ at $c/c^* = 0.176$: ${\color{red}{\diamond}}$  $\dot{\Gamma} = 0.1$, $\color{green}{\star}$ $\dot{\Gamma} = 0.3$, $\color{magenta}{\Box}$ $\dot{\Gamma} = 0.5$, $\color{blue}{\circ}$ $\dot{\Gamma} = 0.7$, $\color{black}{\times}$ $\dot{\Gamma} = 1.0$, $\color{cyan}{\triangleright}$ $\dot{\Gamma} = 2.0$, $\color{blue}{\triangleleft}$ $\dot{\Gamma} = 3.0$,  $\color{black}{\ast}$ $\dot{\Gamma} = 5.0$; (c) Variation of $\eta$ with $\chi$ at $c/c^* = 1.0$: ${\color{red}{\diamond}}$  $\dot{\Gamma} = 0.1$, $\color{green}{\star}$ $\dot{\Gamma} = 0.3$, $\color{magenta}{\Box}$ $\dot{\Gamma} = 0.5$, $\color{blue}{\circ}$ $\dot{\Gamma} = 0.7$, $\color{black}{\times}$ $\dot{\Gamma} = 1.0$, $\color{cyan}{\triangleright}$ $\dot{\Gamma} = 3.0$, $\color{blue}{\triangleleft}$ $\dot{\Gamma} = 5.0$.  In these simulations, $N_{b} = 2$, $b = 25$, $z = 1$, and $d^{\star} = 0.93$. }
\end{figure*}

The existence of a critical mixedness parameter in mixed flows of dilute polymer solutions was first demonstrated by \citet{WooShaqfehJCP2003}, who also proposed an explanation for the significant change in behaviour observed in the response of the solution for values of $\chi$ on either side of $\chi_\text{c}$. They argued that when a molecule, which is aligned along the extension axis, undergoes thermal fluctuations, it suffers a tumbling like motion if it is displaced sufficiently by a fluctuation to end up being aligned along the contraction axis. This can only happen if the angle between the extension and contraction axis is not too large. As can be seen from Fig.~(\ref{MixedFlowAxis}) and Eq.~(\ref{eq:betaPMF}), the magnitude of the angle between the axis is determined by $\chi$, since in terms of $\chi$, $\beta = {\cos}^{-1} \left[(1-\chi)/(1+\chi)\right]$. Since $\beta$ increases with increasing $\chi$, the critical value $\chi_\text{c}$ determines when the angle is too large for thermal fluctuations to cause a molecule to hop from being aligned along the extension axis to being aligned along the contraction axis. Shaqfeh and co-workers have also discussed the scaling of $\chi_\text{c}$ with chain length $N_b$~\cite{WooShaqfehJCP2003,HoffmanShaqfehJoR2007}. However, they have not examined the dependence of $\chi_\text{c}$ on $c/c^{*}$, since they confined their attention to dilute solutions.

A close observation of the change of $\eta$ with $\chi$ in Fig.~\ref{SizeVisVar2}~(d), at small values of $\chi$, appears to suggest that the curves for the various values of $W\!i$ cross each other at a unique value of $\chi$. A zoomed in version of the behaviour in this region is displayed in Figs.~\ref{ChiMixPara}~(a) to~(c),  for dilute solutions, and at two non-zero values of $c/c^{*}$, respectively. The existence of a critical mixedness parameter that demarcates a shear dominated from an extension dominated regime is very clearly demonstrated in these figures. Interestingly, at all concentrations, the value of $\chi_\text{c}$ is independent of the flow strength $W\!i$, and the value of $\eta$ is constant, independent of $W\!i$, at $\chi = \chi_\text{c}$. However, the value of $\chi_\text{c}$ appears to decrease weakly with an increase in $c/c^{*}$, from $\chi_\text{c} \approx 0.04$, both for dilute solutions and at $c/c^{*} = 0.176$, to $\chi_\text{c} \approx 0.03$ at $c/c^{*} = 1$. This can be understood as arising from a decrease in the fluctuations of the polymer coil perpendicular to the extension axis, due to a crowding of molecules with increasing concentration. We can anticipate that the influence of concentration will become more significant for  $c/c^{*} > 1$, when polymer coils begin to interact more strongly with each other. However, a more detailed study of changes in the fluctuations in polymer conformations, and the alignment of molecules relative to the extension and contraction axis, with changes in concentration, is required before a more complete understanding of this observation can be achieved.

\begin{figure*}[tbp]
\begin{center}
\resizebox{9.8cm}{!} {\includegraphics{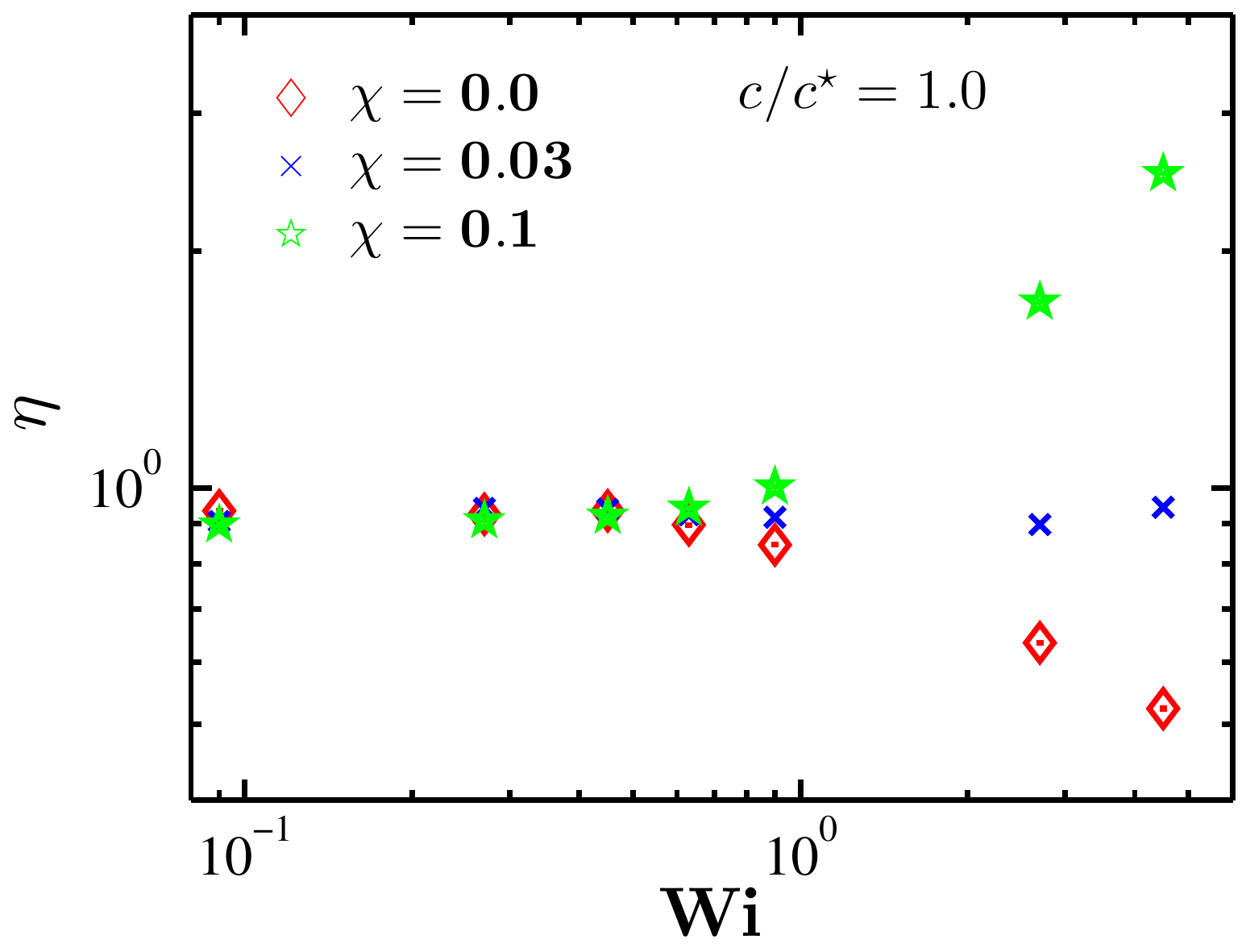}}
\end{center}
\caption{The dependence of viscosity on flow strength at $c/c^{*} = 1$, for three values of the mixedness parameter:  ${\color{red}{\diamond}}$  $\chi = 0$, $\color{green}{\star}$ ${\chi} = 0.1$, and $\color{blue}{\times}$ $\chi_\text{c} = 0.03$.}
\label{etavsGamma}
\end{figure*}

The constancy of $\eta$ with $W\!i$, when $\chi = \chi_\text{c} \approx 0.03$, can be seen more clearly in Fig.~\ref{etavsGamma}, where the dependence of $\eta$ on $W\!i$ is examined, at $c/c^{*} = 1$, for three values of $\chi$, with two of the values ($\chi = 0$ and $\chi = 0.1$), lying on either side of $\chi_\text{c} \approx 0.03$. As discussed earlier, the definition of the viscosity in Eq.~(\ref{eq:EtaH}), ensures that it approaches the Newtonian value for $\dot \gamma \to 0$, and $\dot \epsilon \to 0$. As a result, we expect it to asymptote to the Newtonian value at all values of $\chi$, in the limit of $W\!i \to 0$. This is indeed observed in Fig.~\ref{etavsGamma} as $W\!i \to 0$. At higher values of $W\!i$, the value of $\chi$ determines whether there is shear thinning or extension thickening. At the critical value $\chi_\text{c}$, however, the solution \emph{remains} Newtonian, independent of $W\!i$. This would suggest that in any ensemble of molecules, there are a proportion of molecules undergoing tumbling motions and alignment, and a proportion of molecules undergoing unravelling from coiled to stretched states, such that the net effect is no change of viscosity with increasing deformation rate. Further studies are definitely warranted to verify if this is indeed the case.

\section{Conclusion}
\label{Sec:Conc}

The implementation of periodic boundary conditions for planar mixed flows, in the context of a multi-chain Brownian dynamics simulation algorithm, has been described in some detail. Preliminary results have been obtained on the viscosity of polymer solutions at finite concentrations, when subjected to planar mixed flow. The fascinating behaviour exhibited in these flows, as demonstrated by the various results reported here, has so far not been examined experimentally. In particular, proving the existence of the critical mixedness parameter, and exploring the influence of concentration and chain length in determining its value, would be of great interest. In the context of simulations, determining the scaling of $\chi_c$ with concentration, solvent quality, and chain length, and establishing the correlation between $\chi_c$ and the existence of coil-stretch hysteresis would be extremely valuable. For dilute polymer solutions, it is well known that the size of the coil-stretch hysteresis window observed in planar mixed flows is significantly influenced by the value of the mixedness parameter, vanishing as $\chi \rightarrow 0$, and having a maximum at $\chi \rightarrow 1$.  Studying the dynamics of coil-stretch hysteresis under a variety of circumstances, including varying the concentration, solvent quality and chain length, would provide a fundamental understanding of the hysteresis phenomenon in particular, and of the behaviour of polymer solutions at finite concentrations, in general. The mesoscopic BD algorithm developed in the present work makes it possible to carry out such studies. 

\section*{Acknowledgement}
\noindent This research was supported under Australian Research Council's Discovery Projects funding scheme (project number DP120101322). It was undertaken with the assistance of resources provided at the NCI National Facility systems at the Australian National University through the National Computational Merit Allocation Scheme supported by the Australian Government, and was supported by a Victorian Life Sciences Computation Initiative (VLSCI) grant number VR0010 on its Peak Computing Facility at the University of Melbourne, an initiative of the Victorian Government, Australia.

 \appendix

\section{\label{Sec:Ap} Initial lattice vector for PMFs}

\citet{KraynikReineltIJMF1992} point out in their seminal paper on the derivation of  PBCs for PEF, that ${(\bm{\nabla v})}_{\text{PEF}}$ can be replaced by any diagonalizable constant matrix with real eigenvalues and zero trace. \citet{HuntJCP2010} have exploited this observation by noting that ${(\bm{\nabla v})}_{\text{PMF}}$ is a diagonalizable matrix,
\begin{eqnarray}
\label{eq:diagonalize}
      \begin{pmatrix}
       \dot{\epsilon} & 0 & 0 \\
       \dot{\gamma} & -\dot{\epsilon} & 0  \\
        0 & 0 & 0
    \end{pmatrix}
    &=&
    \begin{pmatrix}
      1 & 0 & 0 \\
      \frac{\dot{\gamma}}{2 \dot{\epsilon}} & 1 & 0 \\
      0 & 0 & 1
    \end{pmatrix}
    \begin{pmatrix}
      \dot{\epsilon} & 0 & 0 \\
      0 & -\dot{\epsilon} & 0 \\
      0 & 0 & 0
    \end{pmatrix} \nonumber \\
    &\times&
    \begin{pmatrix}
      1 & 0 & 0 \\
      -\frac{\dot{\gamma}}{2 \dot{\epsilon}} & 1 & 0 \\
      0 & 0 & 1
    \end{pmatrix} \nonumber 
    = \bm{\Tensor{T}} \cdot \bm{\Tensor{D}} \cdot \bm{\Tensor{T}^{-1}}
  \end{eqnarray}
where $\bm{\Tensor{T}}$ is a transformation matrix that consists of the eigenvectors of ${(\bm{\nabla v})}_{\text{PMF}}$, and the diagonal matrix $\bm{\Tensor{D}}$ has the same component form as ${(\bm{\nabla v})}_{\text{PEF}}$. 
The Kraynik-Reinelt periodic boundary condition for PEF is written in terms of the lattice evolution matrix $\bm{\Tensor{\Lambda}} = \exp{(\bm{\Tensor{D}} t)}$. Similarly for PMF, as the velocity gradient tensor ${(\bm{\nabla v})}_{\text{PMF}}$ is diagonalizable, we can write the lattice evolution matrix $\bm{\Tensor{\Lambda}}'$ as
\begin{eqnarray}
\label{eq:latevolPMF}
\bm{\Tensor{\Lambda}}' &=& \exp{({(\bm{\nabla v})}_{\text{PMF}} t)} = \exp{( \bm{\Tensor{T}} \cdot \bm{\Tensor{D}} \cdot \bm{\Tensor{T}^{-1}} t)}  \nonumber \\
&=&  \bm{\Tensor{T}} \cdot \exp{(\bm{\Tensor{D}} t)} \cdot \bm{\Tensor{T}^{-1}} 
\end{eqnarray}
As ${(\bm{\nabla v})}_{\text{PMF}} = \bm{\Tensor{T}} \cdot \bm{\Tensor{D}} \cdot \bm{\Tensor{T}^{-1}}$ with $\bm{\Tensor{D}}$ being a diagonal matrix, a new set of initial basis vectors,
\begin{equation}
\label{eq:mappingPMF}
\Vector{b_i^0}' = \Vector{b_i^0} \cdot \bm{\Tensor{T}^{-1}} \, \, \, \, \, \, (\text{for} \, \, i = 1, 2, 3)
\end{equation}
exists in PMF, such that this new set is reproducible in the case of PMF \citep{HuntJCP2010}. The tensor ${\bm{\Tensor{T}}}^{-1}$, thus, can be understood as a mapping necessary to make the PEF basis vectors $\Vector{b_i^0}$ (in PEF) reproducible in the PMF (see Refs.~\citenum{KraynikReineltIJMF1992} and~\citenum{JainThesis} for more detail on PEF lattice basis vectors). An equation for the lattice reproducibility condition for PMF can be written as,
\begin{equation}
\label{eq:PMFLatRep1}
\Vector{b_i}' = \Vector{b_i^0}' \cdot  \bm{\Tensor{\Lambda}}'
\end{equation}
where $\Vector{b_i}' $ denotes the lattice vector at time $\tau_p$ (strain period). 
Using this relation, and substituting $\bm{\Tensor{\Lambda}}'$ from Eq. (\ref{eq:latevolPMF}) in Eq. (\ref{eq:PMFLatRep1}) leads to the following simplification
\begin{equation}
\begin{split}
\label{eq:PMFLatRep2}
\Vector{b_i}' (t = \tau_p)  &= \Vector{b_i^0}'  \cdot \bm{\Tensor{\Lambda}}' (\tau_p) \\
   			    &= \Vector{b_i^0} \cdot \bm{\Tensor{T}^{-1}}  \cdot \bm{\Tensor{T}} \cdot \exp{(\bm{\Tensor{D}} t)} \cdot \bm{\Tensor{T}^{-1}}\\
   		            &=    \Vector{b_i^0} \cdot \exp{(\bm{\Tensor{D}} t)} \cdot \bm{\Tensor{T}^{-1}}\\
			    &= \left[N_{i1} \Vector{b_1^0} + N_{i2} \Vector{b_2^0} + N_{i3} \Vector{b_3^0}\right] \cdot  \bm{\Tensor{T}^{-1}} \\
			    &= N_{i1} \Vector{b_1^0} \cdot  \bm{\Tensor{T}^{-1}} + N_{i2} \Vector{b_2^0} \cdot  \bm{\Tensor{T}^{-1}} + N_{i3} \Vector{b_3^0} \cdot  \bm{\Tensor{T}^{-1}} \\
			    &=N_{i1} \Vector{b_1^0}' + N_{i2} \Vector{b_2^0}' + N_{i3} \Vector{b_3^0}' 
\end{split}
\end{equation}  
This equation for the reproducibility condition is identical to the one for PEF \citep{KraynikReineltIJMF1992}, except that $\Vector{b_i^0}$ is replaced by $\Vector{b_i^0}'$. The vectors $\Vector{b_1^0}'$, $\Vector{b_2^0}'$ and $\Vector{b_3^0}'$ can be found easily since $\Vector{b_1^0}$, $\Vector{b_2^0}$ and $\Vector{b_3^0}$ are known for PEF. The mapping of Eq. (\ref{eq:mappingPMF}) is applied to $\Vector{b_i^0}$ to obtain a reproducible lattice under mixed flow as follows.
\begin{equation}
\begin{split}
\label{eq:b10prime}
\Vector{b_1^0}' &= \Vector{b_1^0} \cdot  \bm{\Tensor{T}^{-1}} \\
		& =     \begin{pmatrix}
      \cos \theta & \sin \theta & 0 
    \end{pmatrix}
    \begin{pmatrix}
      1 & 0 & 0 \\
      -\frac{\dot{\gamma}}{2 \dot{\epsilon}} & 1 & 0 \\
      0 & 0 & 1
    \end{pmatrix} \\
    &= \left[\left(\cos \theta - \frac{\dot{\gamma}}{2 \dot{\epsilon}} \sin \theta\right), \sin \theta, 0 \right]
\end{split}
\end{equation} 
\begin{equation}
\begin{split}
\label{eq:b20prime}
\Vector{b_2^0}' &= \Vector{b_2^0} \cdot  \bm{\Tensor{T}^{-1}} \\
		& =     \begin{pmatrix}
      -\sin \theta & \cos \theta & 0 
    \end{pmatrix}
    \begin{pmatrix}
      1 & 0 & 0 \\
      -\frac{\dot{\gamma}}{2 \dot{\epsilon}} & 1 & 0 \\
      0 & 0 & 1
    \end{pmatrix} \\
    &= \left[\left(-\sin \theta - \frac{\dot{\gamma}}{2 \dot{\epsilon}} \cos \theta\right), \cos \theta, 0 \right]
\end{split}
\end{equation} 
\begin{equation}
\begin{split}
\label{eq:b30prime}
\Vector{b_3^0}' &= \Vector{b_3^0} \cdot  \bm{\Tensor{T}^{-1}} \\
		& =     \begin{pmatrix}
      0 & 0 & 1 
    \end{pmatrix}
    \begin{pmatrix}
      1 & 0 & 0 \\
      -\frac{\dot{\gamma}}{2 \dot{\epsilon}} & 1 & 0 \\
      0 & 0 & 1
    \end{pmatrix} \\
    &= \left[0, 0, 1 \right]
\end{split}
\end{equation} 
where $\theta$ is the magic angle, which is similar to that for PEF. In contrast to PEF, where the basis lattice vectors are orthogonal, in the case of PMF, they are non-orthogonal and not equal in length. If the elongational rate is high or the shear rate is small, these lattice vectors becomes almost orthogonal and equal in length. These basis lattice vectors are used as an initial lattice configuration.



\bibliographystyle{elsarticle-num-names}

\bibliography{bibliography1.bib}







\end{document}